\documentclass[11pt]{article}
\usepackage{graphicx,amsmath,bm, amsthm,mathrsfs,amssymb, braket, verbatim}
\usepackage[usenames]{color}
\usepackage{ulem,mathtools}
\usepackage{authblk}
\usepackage{cite}
\usepackage{tcolorbox}
\usepackage[section]{placeins}
\usepackage{placeins}

\newcommand{\ee}{\end{equation}}

\newcommand{\reff}[1]{(\ref{#1})}
\newcommand{\beq}{\begin{equation}}
\newcommand{\eeq}[1]{\label{#1}\end{equation}}
\newcommand{\beqa}{\begin{eqnarray}}
\newcommand{\eea}{\end{eqnarray}}
\newcommand{\eeqa}[1]{\label{#1}\end{eqnarray}}
\newcommand{\beg}{\begin{equation*}}
\newcommand{\eeg}{\end{equation*}}

\newcommand{\bsplit}{\begin{split}}
\newcommand{\esplit}{\end{split}}

\usepackage{subfig} 
\usepackage{circuitikz} 
\usepackage[capposition=bottom]{floatrow} 
\usepackage{epigraph} 
\usepackage{appendix}
\usepackage{rotating}
\allowdisplaybreaks

\title{The non-minimally coupled gravitating vortex:\\ 
phase transition at critical coupling $\xi_c$ in AdS$_3$}

\author[]{Ariel Edery\thanks{aedery@ubishops.ca}}

\affil[]{Department of Physics and Astronomy, Bishop's University, 2600 College Street, Sherbrooke, Qu\'{e}bec, Canada, J1M 1Z7.\vspace{1em}}

\begin{document}
\date{}
\maketitle
\begin{abstract}
We consider the Nielsen-Olesen vortex non-minimally coupled to Einstein gravity with cosmological constant $\Lambda$. A non-minimal coupling term $\xi\,R\,|\phi|^2$ is natural to add to the vortex as it preserves gauge-invariance (here $R$ is the Ricci scalar and $\xi$ a dimensionless coupling constant). This term plays a dual role: it contributes to the potential of the scalar field and to the Einstein-Hilbert term for gravity. As a consequence, the vacuum expectation value (VEV) of the scalar field and the cosmological constant in the AdS$_3$ background depend on $\xi$. This leads to a novel feature: there is a critical coupling $\xi_c$ where the VEV is zero for  
$\xi\ge \xi_c$ but becomes non-zero when $\xi$ crosses below $\xi_c$ and the gauge symmetry is spontaneously broken. Moreover, we show that the VEV near the critical coupling has a power law behaviour proportional to $|\xi-\xi_c|^{1/2}$. Therefore 
$\xi_c$ can be viewed as the analog of the critical temperature $T_c$ in Ginzburg-Landau (GL) mean-field theory where a second-order phase transition occurs below $T_c$ and the order parameter has a similar power law behaviour proportional to $|T-T_c|^{1/2}$ near $T_c$. The plot of the VEV as a function of $\xi$ shows a clear discontinuity in the slope at $\xi_c$ and looks similar to plots of the order parameter versus temperature in GL theory. The critical coupling exists only in an AdS$_3$ background; it does not exist in asymptotically flat spacetime (topologically a cone) where the VEV remains at a fixed non-zero value independent of $\xi$. However, the deficit angle of the asymptotic conical spacetime depends on $\xi$ and is no longer determined solely by the mass; remarkably, a higher mass does not necessarily yield a higher deficit angle. The equations of motion are more complicated with the non-minimal coupling term present. However, via a convenient substitution one can reduce the number of equations and solve them numerically to obtain exact vortex solutions. 
\end{abstract}
\setcounter{page}{1}
\newpage
\section{Introduction}\label{Intro}

In this work, we consider the Nielsen-Olesen vortex, a $2+1$-dimensional abelian Higgs model, non-minimally coupled to Einstein gravity with and without cosmological constant. Compared to previous work on the effects of gravity on vortices \cite{Cadoni, Rich, Ariel, Albert} the new ingredient in the action is the non-minimal coupling term $\xi\,R\,|\phi|^2$ where $R$ is the Ricci scalar, $\xi$ is a dimensionless coupling constant and $\phi$ is a complex scalar field. When gravity is present, it is perfectly fitting to add this term to the action as it preserves the local $U(1)$ gauge invariance of the vortex.  

The non-minimal coupling term changes the physical landscape significantly, in a qualitative fashion. This is related to the dual role that it plays: it acts as part of the potential for the scalar field but also contributes to the Einstein-Hilbert term for gravity. As a consequence, the old parameters when $\xi=0$ such as the VEV $v$, cosmological constant $\Lambda$ and $\alpha$ (proportional to the inverse of Newton's constant)  become effectively the VEV $v_{eff}$, the asymptotic cosmological constant $\Lambda_{eff}$ and $\alpha_{eff}$ respectively that now depend on the coupling $\xi$. The novel feature that emerges is that in an AdS$_3$ background, where $\Lambda_{eff}$ is non-zero and negative, there exists a critical coupling $\xi_c$ where the VEV $v_{eff}$ is zero for $\xi$ at or above $\xi_c$ but is non-zero when $\xi$ crosses below $\xi_c$. When the VEV crosses from zero to non-zero at $\xi_c$, the local $U(1)$ gauge symmetry is spontaneously broken corresponding to a phase transition to a vortex. The critical coupling $\xi_c$ acts like the analog of the critical temperature $T_c$ in Ginzburg-Landau (GL) mean-field theory where the order parameter is zero above $T_c$ but is non-zero below $T_c$ \cite{Justin,Annett}. There is a second-order phase transition when the temperature crosses below $T_c$ and this is typically accompanied by a symmetry that is spontaneously broken. The analogy between $\xi_c$ and $T_c$ can be made quantitative. Near $\xi_c$, we show that the VEV $v_{eff}$ has a power-law behaviour proportional to $|\xi-\xi_c|^{1/2}$ which is similar to the $|T-T_c|^{1/2}$ power-law behaviour of the order parameter near $T_c$ in GL mean-field theory \cite{Justin,Annett}; both have a critical exponent of $1/2$. The plot of the VEV versus the coupling $\xi$ looks very similar to the plot of the order parameter versus temperature $T$ in GL mean-field theory and in both cases there is a discontinuity in the slope at the critical point where the slope diverges. 

The magnitude of the scalar field, represented by the function $f(r)$, starts at zero at the origin $r=0$ and reaches its VEV asymptotically (at a large radius, the computational boundary $R$ which represents formally infinity). An important feature is that the scalar field reaches its VEV slower, over a larger radius, as one approaches the critical coupling $\xi_c$. In other words, the core of the vortex extends out further. The plot of the scalar field's ``extension" \footnote{The extension is defined here as the radius where it reaches $99.9\%$ of its VEV.} as a function of $\xi$ shows a dramatic increase near the critical coupling $\xi_c$. We show analytically that the extension is expected to diverge in the limit $\xi \to \xi_c$. This is the analog to the divergence of the coherence length at the critical temperature $T_c$ in GL mean-field theory \cite{Justin,Annett}. We also plot the extension of the magnetic field which shows a similar trend; starting at its peak value at the origin, it falls off slower (extends further out) as one approaches the critical coupling $\xi_c$.      

We derive analytical expressions for the VEV $v_{eff}$ and the asymptotic cosmological constant $\Lambda_{eff}$ as a function of $\xi$ and four other parameters that appear in the Lagrangian. When $\xi=0$, $v_{eff}$ reduces to $v$ and $\Lambda_{eff}$ reduces to $\Lambda$. However, when $\xi \ne 0$, $v_{eff}$ does not depend only on $v$ and $\xi$ and $\Lambda_{eff}$ does not depend only on $\Lambda$ and $\xi$. They depend each on five parameters in total. A non-zero $\xi$ therefore causes $v_{eff}$ and $\Lambda_{eff}$ to have a dependence on extra parameters besides itself compared to $\xi=0$. This wider influence ultimately stems from the aforementioned dual role that the non-minimal coupling term plays. 

An important point is that the critical coupling exists only in asymptotic AdS$_3$ spacetime; it does not exist in asymptotically flat spacetime ($\Lambda_{eff}=0$) where the VEV is a fixed non-zero constant independent of $\xi$. However, the non-minimal coupling term still plays a significant role in a flat background. In $2+1$-dimensional General Relativity without cosmological constant, it is well known that outside matter the spacetime is locally flat but has the topology of a cone whose deficit angle is proportional to the mass \cite{Deser}. However, we found that the deficit angle was not determined solely by the mass of the vortex but also depended on the coupling $\xi$. One remarkable consequence of this is that a higher mass did not necessarily yield a higher deficit angle. 

The focus of this paper is to study how the vortex changes with the coupling $\xi$.     The effect of other parameters such as $\Lambda$, $v$ and the winding number $n$ has already been studied in previous work \cite{Ariel}. We therefore fix all other parameters and obtain numerical results for different values of $\xi$. With the non-minimal coupling term, the equations of motion are more complicated. Nonetheless, via a convenient substitution, one can reduce the number of equations and solve them numerically. In an AdS$_3$ background, we obtained vortex solutions for nine values of the coupling $\xi$. These ranged from $-0.14$ to $0.095$ (near $\xi_c$) and included the case $\xi=0$. For the parameters chosen, the critical coupling turned out to be equal to $\xi_c=2/21\approx 0.0952$. Note that $\xi_c$ is an upper bound as the VEV is zero for any $\xi$ above this value. For each $\xi$, we provide plots of the scalar field $f(r)$, gauge field $a(r)$, metric field $A(r)$ and magnetic field $B_m(r)$. In a table, for each $\xi$, we state the numerical values obtained for the VEV $v_{eff}$, the cosmological constant $\Lambda_{eff}$, the ADM mass, the peak value of the magnetic field and the numerically integrated magnetic flux. The expected theoretical values for $v_{eff}$ and $\Lambda_{eff}$ obtained from our derived analytical expressions are also quoted in the table. The numerical values and the theoretical expectations for the VEV, cosmological constant and magnetic flux, matched almost exactly (to great accuracy, within three or four decimal places). This provides a strong mutual confirmation of both our numerical simulation and our derived analytical expressions.  We verified numerically that the VEV near $\xi_c$ indeed obeys the power law $|\xi-\xi_c|^{1/2}$. As previously mentioned, the critical exponent of $1/2$ points to a clear analogy with GL mean-field theory where $\xi_c$ acts as the analog of the critical temperature $T_c$. For asymptotically flat spacetime, we considered five values of $\xi$ ranging from $-0.4$ to $+0.4$. The metric field $A(r)$ starts at unity at the origin $r=0$ but then dips below unity and reaches asymptotically (at sufficiently large radius) a plateau at a positive constant value (labelled $D$) that is different for each $\xi$. This is in stark contrast to AdS$_3$ where the metric field $A(r)$ grows as $r^2$ at large radius. The mass and the deficit angle at each $\xi$ are calculated from the numerical value obtained for $D$.   

We now place this paper in context, with a focus on previous studies of gravitating vortices that we referred to earlier \cite{Cadoni, Rich, Ariel, Albert}. It was recognized a long time ago that Einstein gravity in $2+1$ dimensions yields a locally flat spacetime outside localized sources, albeit with the topology of a cone \cite{Deser}. However, things become interesting when one includes a negative cosmological constant as this leads to the famous BTZ black holes \cite{BTZ1,BTZ2}. Later, in a higher-derivative extension of Einstein gravity in $2+1$ dimensions called Bergshoeff-Hohm-Townsend (BHT) massive gravity \cite{BHT1}, black hole solutions in both de Sitter and anti-de Sitter space were found as well as wormhole solutions, kinks, and gravitational solitons \cite{Oliva}. An analytical study of black holes with spherical scalar hair in AdS$_3$ was then later studied \cite{Cadoni}. Closer to our topic of interest, they also constructed black hole vortex solutions with a complex scalar field. These solutions departed from the conventional non-singular vortex in two ways. The scalar field had a singularity at the origin and asymptotically tended towards zero which satisfied the Breitenlohner-Freedman bound \cite{BF} in AdS$_3$ but was not the minimum of the potential. In \cite{Rich}, how vortices affect the tunneling decay of a false vacuum under Einstein gravity was studied and it was found that compared to Coleman-de Luccia bubbles \cite{Coleman} the tunneling exponent was less by a factor of a half. Hence vortices are short-lived and become of cosmological interest \cite{Rich}. The non-singular vortex under Einstein gravity in an AdS$_3$ and Minkowski background was first studied in \cite{Ariel}. These were not black hole solutions as in \cite{Cadoni}. Non-singular vortex solutions were found numerically for different values of the cosmological constant $\Lambda$, VEV $v$ and winding number $n$. Two expressions for the (ADM) mass of the vortex were obtained: one in terms of the metric and one as an integral overly purely matter fields. The latter showed that the mass was roughly proportional to $n^2\,v^2$ (an $n^2$ dependence had also been found in \cite{Cadoni}). The mass of the vortex increased as the magnitude of the cosmological constant increased and led to a slightly smaller core for the vortex. Later, work was then extended to include singular vortex solutions besides non-singular ones \cite{Albert}. Vortices with conical singularities were obtained in flat backgrounds and BTZ black hole solutions were obtained in curved backgrounds, though it was found that the vortex cannot ultimately hold a black hole at its core \cite{Albert}. Our present paper introduces the non-minimal coupling term which is missing in all previous studies of gravitating vortices. As previously pointed out, this term preserves the local $U(1)$ gauge invariance of the vortex and is therefore a perfectly natural candidate to add to the action when gravity is present. We already discussed how this term changes the physics significantly, qualitatively.    

Our paper is organized in the following fashion. In section 2, we obtain analytical expressions for the VEV $v_{eff}$ and the cosmological constant $\Lambda_{eff}$ in terms of $\xi$ and other parameters. Details of the derivation are relegated to Appendix A. We also obtain an expression for the critical coupling $\xi_c$ in terms of the parameters of the theory and discuss the analogy with the critical temperature $T_c$ in GL mean-field theory. In section 3 we state the equations of motion in an abbreviated form and in Appendix B we write down the full equations that are used in our numerical simulation. In section 4 we obtain analytical expressions for the asymptotic metric. In section 5 we obtain an expression for the ADM mass and also obtain an expression for the deficit angle in asymptotically flat space. In section 6 we state the expression for the magnetic field and derive a formula for the magnetic flux which is a topological invariant independent of $\xi$. In section 7 we present all our numerical results in plots and tables for different values of the coupling $\xi$ in both an AdS$_3$ and Minkowski background. Before presenting the numerical results, we obtain useful analytical expressions for the behaviour of the scalar, gauge and metric field asymptotically and near the origin. We end with our conclusion in 
section 8 where among other things, we discuss an interesting and challenging problem to solve in the future.                 

\section{Lagrangian for the vortex non-minimally coupled to Einstein gravity}
The vortex non-minimally coupled to Einstein gravity with cosmological constant has the following Lagrangian density in $2+1$ dimensions:
\beq
\mathcal{L} = \sqrt{-g}\Big(\alpha\,(R-2 \Lambda) -\dfrac{1}{4} F_{\mu\nu}F^{\mu\nu} -\dfrac{1}{2}(D_{\mu} \phi)^{\dagger}(D^{\mu} \phi) +\xi\, R\,|\phi|^2-\dfrac{\lambda}{4}(|\phi|^2-v^2)^2\Big)\,.
\eeq{LDensity}
Here $\phi$ is a complex scalar field, $F_{\mu\nu}$ is the usual electromagnetic field tensor, $R$ is the Ricci scalar, $\Lambda$ is a cosmological constant, the constant $\alpha$ is equal to $\frac{1}{16 \pi G}$ where $G$ is Newton's constant and $\xi$ is a dimensionless coupling constant. The interaction with the gauge field $A_{\mu}$ is incorporated via the usual covariant derivative $D_{\mu} \phi=\partial_{\mu} \phi +i\, e A_{\mu} \phi$ where $e$ is a coupling constant. The constants $\lambda$ and $v$ are parameters that enter into the potential for the scalar field. The constants $\alpha$, $\lambda$ and $v$ are positive whereas $\xi$ can be positive, negative or zero. In $2+1$-dimensional General Relativity, positive $\Lambda$ do not yield black holes (i.e. the famous BTZ black holes require negative $\Lambda$). Similarly, positive $\Lambda$ do not support vortices \cite{Ariel} and the non-minimal coupling term does not change that fact. We will see that $\Lambda$ must be either negative or zero which will ultimately yield asymptotic AdS$_3$ or Minkowski spacetime respectively.   

The Lagrangian density has a local $U(1)$ symmetry; it is invariant under the following gauge transformations
\begin{align}
&\phi(x) \to e^{i\,e\,\eta(x)}\,\phi(x)\\
&A_{\mu}(x) \to A_{\mu}(x)-\partial_{\mu}\eta(x)
\label{Trans}
\end{align}
where $\eta(x)$ is an arbitrary function. The non-minimal coupling term $\xi\, R\,|\phi|^2$ is clearly invariant under the above gauge transformation and is therefore a perfectly natural ingredient to add to the gravitating vortex.

\subsection{The VEV and cosmological constant as a function of $\xi$}
When $\xi=0$, the VEV and cosmological constant are simply $v$ and $\Lambda$ respectively. When $\xi \ne 0$, the VEV and cosmological constant change and become functions of $\xi$ and other parameters. These will be labeled by $v_{eff}$ and $\Lambda_{eff}$ to denote that they are the actual (effective) VEV and cosmological constant respectively for general coupling $\xi$. In this section we determine expressions for them. This requires one to know only the asymptotic behaviour of the fields and this can be determined directly from the Lagrangian without working out the full equations of motion.   

Asymptotically, one reaches the vacuum when the asymptotic spacetime is either AdS$_3$ or Minkowski; these are maximally symmetric spacetimes that can be viewed as the ground states of General Relativity \cite{Carroll}. In this asymptotic region, the kinetic term for the scalar field and gauge field tend to zero: $-\frac{1}{2}(D_{\mu} \phi)^{\dagger}(D^{\mu} \phi)\to 0$ and $-\frac{1}{4} F_{\mu\nu}F^{\mu\nu}\to 0$. This occurs when asymptotically the magnitude of the scalar field approaches the minimum of the potential (the non-zero VEV) and the gauge field approaches a non-zero constant equal to the winding number $n$. In $2+1$ dimensions, the asymptotic value of the Ricci scalar is given by\footnote{Note that the vacuum Einstein field equations with cosmological constant $\Lambda_{eff}$ yield $R=4 \Lambda_{eff}$ in $3+1$ dimensions but $R=6 \Lambda_{eff}$ in $2+1$ dimensions.} $6\, \Lambda_{eff}\,$  where $\Lambda_{eff}$ is either negative (AdS$_3$ background) or zero (Minkowski background). The potential for the scalar field can be readily picked out from the Lagrangian and asymptotically is given by 
\begin{align}
V(|\phi|)=\dfrac{\lambda}{4}(|\phi|^2-v^2)^2-\xi\, R\,|\phi|^2=\dfrac{\lambda}{4}(|\phi|^2-v^2)^2 - 6\,\xi\,\Lambda_{eff} |\phi|^2\,.
\label{Potential}
\end{align}
The VEV occurs at the minimum of this potential where the derivative with respect to $|\phi|$ is zero. This yields two possibilities: $|\phi|=0$ and the solution
\begin{align}
|\phi|^2=v_{eff}^2= v^2 + \dfrac{12 \,\xi \,\Lambda_{eff}}{\lambda}\,.
\label{Vev1}
\end{align} 
When $v_{eff}^2$ is positive, $v_{eff}$ is the minimum of the potential and corresponds to the VEV (and $|\phi|=0$ is a local maximum).  In this case, since the VEV is non-zero, the gauge symmetry is spontaneously broken. When $v_{eff}^2$ is negative (and hence $v_{eff}$ is imaginary), this signals that $|\phi|=0$ is now the minimum of the potential (the VEV). A zero VEV corresponds to the unbroken phase. 

With the non-minimal coupling term $\xi \, R \,\phi^2$ term present in the action, the cosmological constant asymptotically is no longer $\Lambda$ but $\Lambda_{eff}$; this is governed by the equation 
\begin{align}
\alpha (R-2\,\Lambda) +\xi \,R\, v_{eff}^2- \dfrac{\lambda}{4}(v_{eff}^2-v^2)^2=(\alpha +  v_{eff}^2\,\xi)(R- 2\,\Lambda_{eff})\,.
\label{Cosmo1}
\end{align} 
If we substitute $R=6 \,\Lambda_{eff}$ above, we can solve the two equations \reff{Vev1} and \reff{Cosmo1} for $v_{eff}$ and $\Lambda_{eff}$ as a function of $\xi$ and the other parameters of the theory. This is worked out in Appendix A and the equations are \reff{Veff4} and \reff{Leff4}:   
\begin{align}
v_{eff}=\Bigg[2 v^2 + \frac{\alpha}{\xi} - \frac{\sqrt{(\alpha +v^2\,\xi)^2 - 24 \,\alpha\,\Lambda\,\xi^2/\lambda}}{\xi}\,\Bigg]^{1/2}
\label{Veff}
\end{align}
and 
\begin{align}
\Lambda_{eff}=\dfrac{\lambda}{12\,\xi^2}\Big(\alpha + v^2\,\xi -\sqrt{
(\alpha +v^2\,\xi)^2-24\,\alpha\,\Lambda\,\xi^2/\lambda}\,\Big)\,.
\label{Leff}
\end{align}
Equation \reff{Cosmo1} also implies that the coefficient in front of $R$ asymptotically is not $\alpha$ but
\begin{align}
\alpha_{eff}=\alpha + v_{eff}^2\,\xi\,.
\label{alpha_eff}
\end{align}
Newton's constant asymptotically is obtained from $\alpha_{eff}$ so that 
the condition $\alpha_{eff}>0$ must be satisfied. We expect that $\lim_{\,\xi \to 0} v_{eff}=v$, $\lim_{\,\xi \to 0} \Lambda_{eff}=\Lambda$ and $\lim_{\,\xi \to 0} \alpha_{eff}=\alpha$; this is in fact the case as one can readily check.  When $\Lambda$ in \reff{Leff} is negative, this yields a negative $\Lambda_{eff}$ so that the background is AdS$_3$. In that case, $v_{eff}$ and $\Lambda_{eff}$ change with $\xi$. However, when $\Lambda=0$ and $\alpha + v^2\,\xi\,>\,0$ one obtains $\Lambda_{eff}=0$ and $v_{eff}=v$ regardless of the value of $\xi$ or the other parameters. Therefore, in a Minkowski background ($\Lambda_{eff}=0$) the VEV remains constant at $v$ as $\xi$ changes. Note that $\Lambda=0$ with $\alpha +v^2 \,\xi<0$ is not a physically viable option as it leads to a negative 
$\alpha_{eff}$ i.e. one obtains $v_{eff}^2= 3 \,v^2  + \frac{2 \,\alpha}{\xi}$ so that 
$\alpha_{eff}=\alpha + v_{eff}^2 \,\xi$ is equal to $3\,(\alpha +v^2\,\xi)$ which is negative.      

When $\xi=0$, $v_{eff}$ is simply $v$ but when $\xi\ne 0$, $v_{eff}$ does not depend only on $v$, $\xi$ and $\lambda$ but also on the gravitational parameters $\alpha$ and $\Lambda$. Similarly, when $\xi\ne 0$, $\Lambda_{eff}$ does not depend only on 
$\Lambda$, $\xi$ and $\alpha$ but also on the parameters $v$ and $\lambda$ appearing in the scalar potential. We see that the non-minimal coupling term has a wide reach because of the dual role it plays in affecting simultaneously the potential of the scalar field and the Einstein-Hilbert gravitational term. 

 \subsection{Critical coupling $\xi_c$}
The VEV, given by \reff{Vev1}, is equal to zero at a critical coupling $\xi_c$. This occurs when   
\begin{align}
2 v^2 + \frac{\alpha}{\xi} - \frac{\sqrt{(\alpha +v^2\,\xi)^2 - 24 \,\alpha\,\Lambda\,\xi^2/\lambda}}{\xi}=0
\label{effective}
\end{align}
which has the solution
\begin{align}
\xi_c=-\dfrac{2\,v^2 \,\alpha\,\lambda}{3\,(v^4 \lambda + 8\,\alpha\,\Lambda)}
\label{Critical}
\end{align}
if the condition $\alpha + 2 \,v^2\,\xi>0$ is satisfied. This condition implies that $v^4 \lambda + 8\,\alpha\,\Lambda$ in the denominator of \reff{Critical} is negative. The critical coupling is therefore positive and exists only when $\Lambda$ is negative and obeys the inequality $\Lambda<- \frac{v^4 \lambda}{8\,\alpha}$. A negative $\Lambda$ implies $\Lambda_{eff}<0$ so that the spacetime is asymptotically AdS$_3$. In particular, the case $\Lambda=0$ (which yields $\Lambda_{eff}=0$) has no critical coupling and has a fixed VEV at $v$. There is therefore no critical coupling in asymptotic Minkowski spacetime. The critical coupling exists only in AdS$_3$ when $\Lambda<- \frac{v^4 \lambda}{8\,\alpha}$. What happens when $\Lambda$ is negative but falls in the range $- \frac{v^4 \lambda}{8\,\alpha}<\Lambda <0$? The spacetime is asymptotically AdS$_3$ since $\Lambda_{eff}<0$ and the VEV changes with $\xi$ but it always remains above zero; there is no transition from the unbroken phase (zero VEV) to the broken phase (non-zero VEV). Note that the value of the critical coupling does not depend on the winding number $n$.

When the critical coupling exists, the VEV is zero for $\xi\ge \xi_c$, but is non-zero and grows as $\xi$ decreases below $\xi_c$. A phase transition from a symmetric (unbroken) phase to a spontaneously broken phase occurs when $\xi$ crosses below $\xi_c$. In figure 1 below,  we plot $v_{eff}$ as a function of $\xi$ (for parameters $\alpha=1$, $v=1$, $\lambda=1$ and $\Lambda=-1$). Since $\Lambda<-\frac{v^4 \lambda}{8\,\alpha}=-1/8$, the condition for a critical coupling is satisfied and its value from \reff{Critical} is $\xi_c= 2/21 = 0.0952$. We see that the VEV is zero above $\xi_c=0.0952$ but becomes non-zero and increases as $\xi$ decreases below $\xi_c$. The VEV is continuous but one can readily see that the derivative (slope of graph) is discontinuous at $\xi_c$. We will see that in fact the slope diverges at that point.  

\begin{figure}[t]
	\centering
		\includegraphics[scale=1.0]{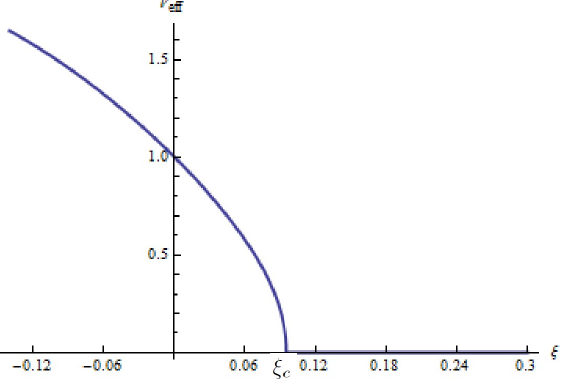}
		\caption{The VEV $v_{eff}$ as a function of $\xi$ plotted for parameters $\alpha=1$, $v=1$, $\lambda=1$ and $\Lambda=-1$. The VEV is zero at or above $\xi_c=0.0952$ and transitions to a non-zero value below $\xi_c$ where it increases as $\xi$ decreases. When $\xi$ crosses below $\xi_c$, there is a transition from a symmetric phase to a spontaneously broken phase. Note that, as expected, the VEV is equal to $v=1$ at $\xi=0$.}
	\label{VEV}
\end{figure}

Figure 1 should bring to mind the graph (see fig. 2)\footnote{Image courtesy of C. Lygouras, ``Critical behavior of the order parameter and specific heat in the second-order phase transition from Landau theory", May 4, 2020. Wikimedia Commons contributors, `File:LandauTheoryTransitions.svg`, Wikimedia Commons, the free media repository.} of the order parameter as a function of temperature in the Ginzburg-Landau (GL) mean-field theory of second-order phase transitions where the order parameter is zero above a critical temperature $T_c$ but increases above zero below $T_c$. Our critical coupling $\xi_c$ is the analog to the critical temperature $T_c$. We can make this connection more quantitative. In GL mean-field theory, at temperatures $T$ below and near $T_c$, the order parameter is proportional to $(T_c-T)^{1/2}\,$\cite{Justin,Annett} a power law behaviour with critical exponent of $1/2$. The VEV for $\xi$ below and near $\xi_c$ has a similar behaviour. Using \reff{Critical}, we can express $\Lambda$ in terms of $\xi_c$ and substitute this into \reff{Veff} to obtain
\begin{align}
v_{eff}=\Bigg[\,2 v^2+\frac{\alpha }{\xi }-\frac{\sqrt{\alpha^2+2 \,v^2\, \alpha \,\xi + v^4 \,\xi^2-\frac{\xi^2 \,\left(-2 \,v^2 \,\alpha \,\lambda -3 \,v^4 \,\lambda \,\xi_c\right)}{\lambda \,\xi_c}}}{\xi}\,\Bigg]^{1/2}\,.
\end{align}
Expanding $v_{eff}$ above about the critical coupling $\xi_c$ yields 
\begin{align}
v_{eff}= k\,(\xi_c-\xi)^{1/2} + \mathcal{O}\big((\xi_c-\xi)^{3/2}\big)
\label{expansion}
\end{align}
where the proportionality constant is $k=\frac{v}{\sqrt{\xi_c + (2\,v^2\,\xi_c^2)/\alpha}}$. We therefore see that the power law behaviour of the VEV near $\xi_c$ and of the order parameter near $T_c$ in GL theory are similar and have the same critical exponent of $1/2$. From \reff{expansion}, one can readily see that the slope in figure 1 diverges at $\xi_c$ (just like the slope in figure 2 diverges at $T_c$). We will set that the VEV for values of $\xi$ near $\xi_c$ in our numerical simulation follows closely the power law behaviour given by \reff{expansion}.

\begin{figure}[t]
	\centering
		\includegraphics[scale=1.0]{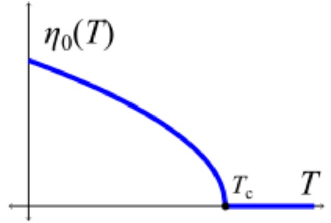}
		\caption{The order parameter $\eta_0(T)$ as a function of temperature in the GL mean-field theory. The order parameter is zero at or above the critical temperature $T_c$ but is non-zero below $T_c$. There is a discontinuity in the slope at $T_c$ and there is a second-order phase transition when the temperature crosses below $T_c$.}                            
	\label{Order}
\end{figure}

We will now determine the equations of motion, solve them numerically and obtain plots of various quantities for different values of the coupling $\xi$. The equations \reff{Veff} and \reff{Leff} for the VEV and cosmological constant that we derived in this section will be used to predict the asymptotic values of our plots and we will see that they match exactly. This provides a strong confirmation of both our derived theoretical results of this section and of our numerical vortex solutions in later sections.

\section{Rotationally symmetric ansatz and the equations of motion}
For the vortex, we consider rotationally symmetric static solutions. The ansatz for the  gauge and scalar field is 
\begin{align}
&A_j(\mathbf{x})=\epsilon_{jk}\hat{x}^k \dfrac{a(r)}{er}\\
&\phi(\mathbf{x}) = f(r)\,e^{i n \,\theta}
\label{ansatz}
\end{align} 
where $a(r)$ and $f(r)$ are functions of $r$ that represent the gauge and scalar field 
respectively. The non-negative integer $n$ is called the winding number.  A $2+1$ dimensional metric which is rotationally symmetric can be expressed as  
\beq
ds^2=- B(r) \,dt^2 +\dfrac{1}{A(r)} \,dr^2 + r^2\, d\theta^2
\eeq{metric} where $A(r)$ and $B(r)$ represent two metric functions of $r$. 

With the ansatz \reff{ansatz} and \reff{metric}, the Langrangian density \reff{LDensity} reduces to 
\beq
\mathcal{L}=\sqrt{B/A}\,r\,\Big(\alpha  \,(R-2\Lambda )-\frac{A (a')^2}{2\, e^2 \,r^2}-\frac{(f')^2 A}{2}- \frac{(n-a)^2\,f^2}{2 \,r^2} +\xi R\,f^2-\frac{\lambda}{4}\,(f^2 -v^2)^2 \Big)\,.
\eeq{LDensity2}
Since $f$ approaches a non-zero constant asymptotically, one requires that $a\to n$ asymptotically (which yields $(n-a)^2\,f^2 \to 0$) so that one avoids a logarithmic divergence in the energy of the vortex \cite{Weinberg,Ariel}. The Ricci scalar is a function of $A$ and $B$ and their derivatives:
\beq
R= \frac{ (B')^2 A}{ 2 B^2}-\frac{ A'}{r }-\frac{A'B'}{2 B}- \frac{B' A}{r B} -\frac{ B'' A}{B}\,.
\eeq{Ricci}
Note that when the complex scalar field is inserted in the Lagrangian density, the winding number $n$ appears but not the coordinate $\theta$ since the phase cancels out. 
The Lagrangian density therefore depends on $r$ only and solutions are rotationally symmetric. The Euler-Lagrange equations of motion for $A(r)$, $B(r)$, $f(r)$ and $a(r)$ are respectively 
\begin{align}
&4\, e^2 \,r \,A \,B' (\alpha +\xi \,f^2+ 2 \,r\, \xi\, f \,f')+B\Big(e^2 r^2 (v^4 \lambda +8 \alpha\, \Lambda)+2 e^2 (n^2-r^2 v^2 \lambda-2 \,n \,a+a^2) f^2
\nonumber\\&\quad\quad+e^2 r^2 \lambda f^4+16 \,e^2 r \,\xi \,A \,f\, f'-2 A (a'^{\,2}+e^2 r^2 f'^{\,2})\Big)=0 \label{EOMA}\\\nonumber\\
&e^2 r^2 \lambda f^4+e^2 r \,(r v^4 \lambda+8 r \alpha \Lambda+4 \alpha A')+2 e^2 f^2 (n^2-r^2 v^2 \lambda-2\, n \,a+a^2+2 r \xi A')\nonumber\\&\quad\quad+2 A \,(a'^{\,2}+e^2 r^2 (1+8 \xi) f'^{\,2})+8 e^2 r \xi f \,\big(r A' f'+2 A \,(f'+r f'')\big)=0\label{EOMB}\\\nonumber\\
&2 r^2 \xi A f B'^{\,2}+r B \,\big(-2 r \xi f A' B'+A\,(r B' f'-4 \xi f (B'+r B''))\big)+B^2 \Big(-2 r^2 \lambda f^3\nonumber\\&\quad\quad-2 f \,(n^2-r^2 v^2 \lambda-2 n a+a^2+2 r \xi A')+r \,(r A' f'+2 A (f'+r f''))\Big)=0 \label{EOMf}\\\nonumber\\
&r A a' B'+B \Big(2 \,e^2 r\, (n-a) f^2-2 A a'+r a' A'+2 \,r A a''\Big)=0\,.\label{EOMa}
\end{align}
We can reduce the above four equations of motion to three by extracting $W(r)=B'/B$
from equation \reff{EOMA} and substituting it into equations \reff{EOMf} and \reff{EOMa}. 
The function $W(r)$ contains $A,f$ and $a$ and their derivatives. The main point is that the three remaining equations no longer have any dependence on $B(r)$. However, the equations become longer especially the one for the function $f(r)$. We write them out in full in Appendix B; equations \reff{EOMB4}, \reff{EOMf4} and \reff{EOMa4} are the equations we solve numerically. To avoid writing out cumbersome lengthy equations here, the three remaining equations are written below using $W(r)$ and $W'(r)$. Note that we need $W'$ because of the appearance of $B''$ in \reff{EOMf}. In particular, $B''/B = W' + W^2$. The three remaining equations are    
\begin{align}
&e^2 r^2 \lambda f^4+e^2 r \,(r v^4 \lambda+8 r \alpha \Lambda+4 \alpha A')+2 e^2 f^2 (n^2-r^2 v^2 \lambda-2\, n \,a+a^2+2 r \xi A')\nonumber\\&\quad\quad+2 A \,(a'^{\,2}+e^2 r^2 (1+8 \xi) f'^{\,2})+8 e^2 r \xi f \,\big(r A' f'+2 A \,(f'+r f'')\big)=0\label{EOMB2}\\\nonumber\\
&2 \,r^2 \,\xi A \,f\, W^2 -2 r^2 \xi \,f \,A'\, W +A\,r\big(r \,W \,f'-4 \xi f (W+r \,(W'+W^2))\big)-2 \,r^2 \lambda f^3\nonumber\\&\quad\quad-2 \,f \,(n^2-r^2 v^2 \lambda-2\,n \,a+a^2+2 \,r \xi \,A')+r \,\big(r A' f'+2 A \,(f'+r f'')\big)=0\label{EOMf2}\\\nonumber\\
&r \,A\, a' \,W + 2 \,e^2\, r\, (n-a) f^2 - 2 \,A \,a'+ r \,a'\, A' + 2 \,r \,A\, a''=0\,.
\label{EOMa2}\end{align}
When $W(r)$ given by \reff{W} is substituted into the above equations we obtain the full equations \reff{EOMB4}, \reff{EOMf4} and \reff{EOMa4}.   
   
\section{Asymptotic analytical solutions}
One can solve analytically for the metric in vacuum by setting $f=v_{eff}$ and 
$a=n$ identically in Eq. \reff{EOMB2}. This yields 
\begin{align}
A'(r)=-r \,\dfrac{(8\,\alpha \,\Lambda + \lambda\,(v_{eff}^2-v^2)^2)}{4(\alpha + \xi\, v_{eff}^2)}
\end{align}
with solution
\beq
A_0(r) = -\dfrac{(8\,\alpha \,\Lambda + \lambda(v_{eff}^2-v^2)^2)}{8(\alpha + \xi\, v_{eff}^2)}\,r^2 + C =- \Lambda_{eff} \,r^2 + C
\eeq{A0}
where the subscript `0' denotes vacuum and $C$ is an integration constant. In the last step we substituted $v_{eff}$ given by \reff{Veff} and this yields $\Lambda_{eff}$ given by \reff{Leff} as the coefficient of $-r^2$ (see also Eq. \reff{Cosmo3}).  Of course, this is exactly what we expect the metric of pure AdS$_3$ to be for cosmological constant $\Lambda_{eff}$. The initial conditions at $r=0$ are determined by the constant $C$. We set $C=1$ since in $2+1$ dimensions this choice avoids a conical singularity at the origin \cite{BTZ1,Deser}. Moreover, $C=1$ also works for the case of vortices embedded in asymptotically Minkowski spacetime ($\Lambda_{eff}=0$). 

We can now solve for the metric function $B_0(r)$ in vacuum by substituting $A_0(r)$ with $C=1$ into Eq.\reff{EOMA}. This yields $B_0(r) = k_0\,(-\Lambda_{eff} \,r^2 + 1)$ where $k_0$ is an integration constant (positive). We can absorb this constant into a redefinition of time in the line element \reff{metric} so that 
\beq
B_0(r)= -\Lambda_{eff}\, r^2 + C = A_0(r)\,.
\eeq{B0}
In the presence of the vortex, we have that $f \to v_{eff}$ and $a \to n$ asymptotically. Note that in contrast to the vacuum case, these are now only the asymptotic values. The vortex departs significantly from that in the core region near the origin. In numerical simulations, $f$ and $a$ start at zero at the origin and reach their asymptotic value (within less than a percent) at a finite radius $R$, the computational boundary which represents formally infinity. The asymptotic form of the metric function $A$ in the presence of matter (the vortex) is obtained again via Eq. \reff{EOMB2} and yields at $r=R$  
\beq
A(R)= -\Lambda_{eff} \,R^2 + D\,. 
\eeq{AR}
The constant $D$ differs from the constant $C$ in \reff{A0}; as one goes through the core of the vortex, one naturally emerges into an asymptotic region that differs from the purely vacuum one and this is reflected in $D$ being a different constant from $C$.  We will see that the (ADM) mass of the vortex will be expressed in terms of $A_0(R)$ and $A(R)$. 

Asymptotically, using \reff{EOMA}, we obtain $B(R)= k \, A(R)$. Here $k$ is an integration constant (positive); it can no longer be absorbed into a redefinition of time since that has been carried out once already with the constant $k_0$. At large radius $R$, in the presence of the vortex, we obtain that $B(R)$ is proportional to $A(R)$ but not equal to it.    
  
\section{Expression for the (ADM) mass of the vortex}   

An important property of a vortex is its finite mass. In curved spacetime, the mass of a localized source is defined as its ADM mass \cite{Poisson}. AdS$_3$ is a maximally symmetric spacetime with isometry group $SO(2,2)$ and has a timelike Killing vector so that a conserved energy (the ADM mass) naturally applies to matter embedded in it. The ADM mass in $2+1$ dimensions can be calculated via the following expression \cite{Poisson}:   
\beq
M= -2 \,\alpha_{eff}\, \lim_{C_t \to R} \oint_{C_t} (k-k_0) \,\sqrt{\sigma} \,N(R) \,d \theta \,.
\eeq{M}
Note that $\alpha_{eff}$, given by \reff{alpha_eff}, must be used here instead of $\alpha$. Here $C_t$ is the circle at spatial infinity where infinity corresponds to the computational boundary $r=R$. The lapse $N(R)$ is given by $\big(B_0(R)\big)^{1/2}=\big(A_0(R)\big)^{1/2}$. The metric on $C_t$ is $\sigma_{AB}$ and $\sqrt{\sigma} =R$ where $\sigma$ is its determinant. The extrinsic curvature of $C_t$ embedded on the two-dimensional spatial surface obtained by setting $t$ to be constant in \reff{metric} is given by $k$ whereas its embedding in the two-dimensional spatial surface of AdS$_3$ is given by $k_0$. A straightfoward calculation yields  
\beq
k= \dfrac{\big(A(R)\big)^{1/2}}{R} \quad;\quad k_0= \dfrac{\big(A_0(R)\big)^{1/2}}{R} 
\eeq{extrinsic}
Substituting all the above quantities into \reff{M} yields our final expression for the ADM mass:
\beq
M= 4 \,\pi \,\alpha_{eff} \,\Big(A_0(R) -[A_0(R)\,A(R)]^{1/2}\Big)\,. 
\eeq{ADM}
We will use the above expression to calculate the ADM mass in an AdS$_3$ background. Note that if  $A(R)=A_0(R)$ one obtains $M=0$ which implies that our definition has set empty AdS$_3$ space to have zero mass. This is the desired and expected result since maximally symmetric spacetimes can be viewed as the ground states of General Relativity \cite{Carroll} and as such are typically set to zero energy.    

The analytical expression \reff{A0} for the vacuum metric $A_0(R)$ is $-\Lambda_{eff}\, R^2 + 1$ and this can be readily calculated for any given $R$. From \reff{AR} we have that $A(R)=-\Lambda_{eff}\, R^2 + D$ where $D$ is a constant. This corresponds to the case with matter (the vortex) and it is obtained by solving the equations of motion numerically since we do not know a priori the value of the constant $D$. The mass $M$ of the vortex is then obtained via \reff{ADM}. Though $A_0(R)$ and $A(R)$ both change with $R$, at a large enough $R$, the mass $M$ hardly changes as $R$ increases and the matter fields $f(r)$ and $a(r)$  plateau to their respective asymptotic values of $v_{eff}$ and $n$ respectively. The value of $A(r)$ at $r=0$ is an initial condition. In vacuum, $A(r)$ must reduce to $A_0(r)$ so that their initial conditions at the origin must match. This implies that $A(0)=A_0(0)=C=1$. 

\subsection{ADM mass in asymptotically flat space and angular deficit} 
In asymptotically flat spacetime where $\Lambda=\Lambda_{eff}=0$, the ADM mass formula \reff{ADM} remains valid but simplifies greatly. We have that $A_0(R)=C=1$ and $A(R)=D$ which yields
\begin{align}
M_{flat}= 4 \,\pi \,\alpha_{eff} \,\big(1 -D^{1/2}\big) 
\label{M2}
\end{align} 
where $\alpha_{eff}=\alpha +\xi\,v^2$ since $v_{eff}=v$. Note that $A_0(r)=B_0(r)$ stay constant at unity for all $r$ (this represents the vacuum Minkowski spacetime). In contrast, $A(r)$ is unity at the origin $r=0$ but dips below unity as $r$ increases  until it plateaus to a positive value $D$ at large radius $R$. The value of $D$ is obtained numerically. Recall that localized matter in $2+1$ dimensions yields an asymptotically Minkowski spacetime with an angular deficit \cite{Deser}. Asymptotically, $A(r)=D$ and the spatial part of the metric \reff{metric} becomes $\frac{dr^2}{D} + r^2 d\theta^2$.  If we define $r_0=r/D^{1/2}$ and $\theta_0= D^{1/2}\, \theta$ we obtain a manifestly flat metric $dr_0^2 + r_0^2 \,d\theta_0^2$ but with $\theta_0$ ranging now from $0$ to $2 \pi D^{1/2}$ instead of $2\,\pi$. Since $0<D<1$ there is an angular deficit of 
\begin{align}
\delta= 2 \pi\, (1-D^{1/2})\,.
\label{delta}
\end{align}
Using \reff{M2} with $\alpha_{eff}=1/(16 \pi G_{eff})$ we obtain that $\delta= 8 \pi\, G_{eff}\,M_{flat}$ which is the formula for the angular deficit produced by a mass $M_{flat}$ in $2+1$ Minkowski spacetime \cite{Deser} if $G_{eff}$ replaces $G$ in \cite{Deser}. Asymptotically, the spacetime is locally flat but topologically a cone. There is however no conical singularity at the origin in our case in contrast to the point mass in \cite{Deser}. The spacetime is smooth at the origin since the vortex by construction is an extended non-singular object. In our case, the conical spacetime is only the asymptotic spacetime and does not extend into the core of the vortex. 

In the original work of \cite{Deser}, the only way to change the angular deficit is to  change the mass since $G$ remains constant. In our case, $G_{eff}$ depends on the coupling $\xi$. Therefore as $\xi$ changes, one can encounter a scenario (and one does as our numerical results will show) where a higher mass yields a smaller deficit angle than a smaller mass. This is another instance of how the non-minimal coupling term plays a novel role.

\section{Magnetic flux as a topological invariant independent of coupling $\xi$}

The vortex contains a magnetic field which we label $B_m$. We will see when we plot our numerical results that it has its maximum at the origin and then decreases towards zero outside a core region. The maximum value of the magnetic field at the origin as well as its profile depends on the coupling $\xi$. After we present our numerical results, we will look at the radial extension of the scalar field as a function of $\xi$, a measure of how far the field extends before it reaches close to its plateau value (the VEV). We will see that the radial extension of the scalar field increases significantly as we approach the critical coupling $\xi_c$. This is analogous to the coherence length in GL mean-field theory which diverges near the critical temperature. We have discussed here the radial extension of the scalar field because we will see that the radial extension of the magnetic field as a function of $\xi$ undergoes the same fate and also increases as we approach the critical coupling $\xi_c$. The magnetic field profile therefore provides us with an additional window into how far the core region of the vortex extends.  

An important property of the magnetic field is that even though its profile changes with the coupling $\xi$, the magnetic flux $\Phi$ obtained by integrating the magnetic field over the entire two-dimensional area stays constant (i.e. it is independent of the value of $\xi$). We show here that the magnetic flux depends only on the winding number $n$ and hence is a topological invariant. The quantity $-\frac{A \,(a')^2}{2 \,e^2 \,r^2}$ appearing in the Lagrangian density \reff{LDensity2} stems from the term $-\frac{1}{4} F_{\mu\nu}F^{\mu\nu}$ and hence is identified with $-B_m^2/2$ where $B_m$ is the magnetic field (no electric field is present hence the absence of an $\tfrac{E^2}{2}$ term). It follows that the magnetic field is given by $B_m= \frac{\sqrt{A}\, a'}{e \,r}$ which reduces to the well-known result $a'/(e\,r)$ for the magnetic field in fixed Minkowski spacetime \cite{Weinberg} where $A(r)=1$ identically. 

The magnetic flux $\Phi$, the integral of the magnetic field over the invariant area element, yields
\begin{align}
\Phi=\int d^2x \,\sqrt{\gamma} \,B_m =\int dr \,d\theta \,\Big(\frac{r}{\sqrt{A}}\Big)\,\big(\frac{\sqrt{A}\, a'}{e \,r}\big)= \frac{2\,\pi}{e} \int_{0}^{R} a' \,dr= \frac{2\,\pi}{e}\,(a(R)-a(0))=\frac{2\,\pi\,n}{e}
\label{Flux}
\end{align}
where $\gamma=r^2/A$ is the determinant of the spatial two-metric obtained from \reff{metric} by setting $t$ to be constant. We used the boundary conditions on the function $a(r)$: $a(R)=n$ and $a(0)=0$. Note that the expression for the magnetic flux $\Phi=\frac{2\,\pi \,n}{e}$ is the same in curved space as it is in fixed Minkowski spacetime \cite{Weinberg}. In the next section where we present our numerical results, we will integrate numerically over the area the different magnetic field profiles for different coupling $\xi$ and show that the result is the same independent of the profile and $\xi$.  Besides demonstrating numerically that the magnetic flux is a topological invariant in curved space, it also provides another check on our numerical simulation. The magnetic flux is ``quantized" as it comes in integer steps of $2\pi/e$. This does not stem from any quantization procedure imposed on the fields but from the topology of the vortex which is characterized by its winding number $n$. 
 
\section{Numerical solutions of vortex in curved space}

The three equations of motion \reff{EOMB4}, \reff{EOMf4} and \reff{EOMa4} are solved numerically to obtain non-singular profiles for the scalar field $f(r)$, the gauge field $a(r)$ and the metric function $A(r)$. The initial conditions at the origin $r=0$ are  
\begin{align}
f(0)=0\quad; \quad a(0)=0\quad;\quad  A(0)=1\,.
\label{boundary}
\end{align}
These initial conditions ensure that our vortex solutions are non-singular. Let $R$ be the computational boundary representing formally infinity. We expect that
\begin{align} 
f(R)=v_{eff}\quad;\quad a(R)=n \quad ;\quad A(R)= D - \Lambda_{eff} \,R^2 
\label{Asym}
\end{align}
where $D$ is a constant that is determined only after running the numerical simulation and depends on the matter distribution of the vortex. The quantity $v_{eff}$ is the value where $f(r)$ plateaus at numerically and we will see that it matches very closely our theoretical prediction given by \reff{Veff}. The winding number of the vortex is given by the positive integer $n$ and we will see that $a(r)$ plateaus at that value numerically. The coefficient $\Lambda_{eff}$ in front of $R^2$ in $A(R)$ can be extracted from our numerical simulation by evaluating $-A''(r)/2$ at $r=R$. We will see that it matches very closely our theoretical prediction for the asymptotic value of the cosmological constant given by \reff{Leff}. We obtain the profiles by adjusting $f'(r)$ and $a'(r)$ near the origin until the curves for $f(r)$ and $a(r)$ plateau towards their respective constant values beyond a certain radius (in our numerical simulations they reach their expected constant values to within less than a tenth of a percent at the computational boundary $R$).   

\subsection{Analytical behaviour of the fields near the origin and asymptotically}

The equations of motion are a long complicated set of coupled non-linear differential equations which require a numerical solution. However, before presenting the numerical results, it is instructive to extract some useful analytical information from the equations. In particular, we will determine the analytical behaviour of the fields near the origin and in the asymptotic region. We will see that the asymptotic profile of a vortex is not supported by a positive cosmological constant $\Lambda_{eff}$; it must be either negative (AdS$_3$ background) or zero (Minkowski background). This is similar to the fact that in $2+1$ dimensional General Relativity (GR), a black hole exists for negative cosmological constant (the BTZ black hole \cite{BTZ1,BTZ2}) but not for positive cosmological constant. There is no black hole in a Minkowski background either but in contrast, one can have a vortex in a Minkowski background.

\subsubsection{Behaviour of $A(r)$, $f(r)$ and $a(r)$ near the origin}
The initial conditions on the fields at $r=0$ are $f(0)=0$, $a(0)=0$ and $A(0)=1$. We would like to know the behaviour of these fields in the vicinity of $r=0$. If we linearize \reff{EOMB4} about $A=1$ we obtain $A(r)=1- r^2 (\frac{v^4\,\lambda}{8 \alpha}+ \Lambda)$. This quadratic behaviour implies that its first derivative $A'(r)$ at $r=0$ is always zero regardless of the parameters so that the metric function always starts out flat at the origin. This is what is observed numerically. Linearizing \reff{EOMa4} about $a=0$ yields $a(r)=b\,r^2$ with $b$ a positive constant. We see that $a(r)$ also starts out flat at the origin since $a'(0)=0$. Again, this agrees with our numerical simulation. Linearizing \reff{EOMf4} about $f=0$ yields $f(r)= c\, r^n$ where $n$ is the winding number and $c$ a positive constant. Near the origin, $f'(r)=c\,n r^{n-1}$ so that $f'(0)=c$ for $n=1$ and $f'(0)=0$ for $n>1$. This implies that $f(r)$ starts off flat at the origin when $n>1$ but with a positive slope when $n=1$. Note that the fields near $r=0$ have no dependence on the coupling $\xi$.  

\subsubsection{Behaviour of $A(r)$, $f(r)$ and $a(r)$ asymptotically} 
Asymptotically, the metric function $A(r)$ is given by $D- \Lambda_{eff} \,r^2$ where 
$D$ is a constant. The matter fields $a$ and $f$ plateau to their constant values of $n$ and $v_{eff}$ respectively asymptotically. We would like to know their behavior as they approach these constant values. At large $r$ we can write $a(r)=n-\epsilon(r)$ and $f(r)=v_{eff}-\beta(r)$ where $\epsilon$ and $\beta$ are small positive perturbations which must vanish asymptotically. Substituting these expressions into equation \reff{EOMa4} and \reff{EOMf4} and keeping only terms linear in $\epsilon$ and $\beta$ yields the differential equations
\begin{align}
&e^2 \,v_{eff}^2\, \epsilon(r) + r \,\Lambda_{eff} \,\epsilon'(r)+ r^2 \,\Lambda_{eff} \,\epsilon''(r)=0 \\\nonumber\\
& 2\, v_{eff}^2 \left(\alpha_{eff}\,\lambda -24 \,\Lambda_{eff} \,\xi^2\right) \beta(r) +r\, \Lambda_{eff} \left(\alpha_{eff}+16\,v_{eff}^2 \,\xi^2\right) \left(3 \,\beta'(r) +r \beta''(r)\right)=0\,.
\label{Linear}
\end{align}
The above equations are valid only for the case $\Lambda_{eff}\ne 0$ (the case $\Lambda_{eff}=0$ will be treated separately). Both equations have power law fall off solutions 
\begin{align}
&\epsilon(r)=b\, r^{-\,\dfrac{e\,v_{eff}}{(-\Lambda_{eff})^{1/2}}}\label{eps}\\
&\beta(r)= c\, r^{-1-\Big[\dfrac{-\alpha_{eff}\,\Lambda_{eff} + 2 \,\alpha_{eff}\,
v_{eff}^2 \,\lambda - 64 \,v_{eff}^2 \,\Lambda_{eff} \,\xi^2}{-\alpha_{eff}\,\Lambda_{eff} - 16 \,v_{eff}^2 \,\Lambda_{eff} \,\xi^2}\Big]^{1/2}}
\label{beta}
\end{align}
where $b$ and $c$ are positive constants. Since \reff{eps} is valid only if $\Lambda_{eff}$ is negative, the above profiles apply only to an AdS$_3$ background. An important point is that the profile of a vortex which requires the gauge field $a$ to plateau at $n$ and $f$ to plateau at $v_{eff}$ is not supported by a positive $\Lambda_{eff}$. It is supported by a negative $\Lambda_{eff}$ and as we will now see, also by a zero $\Lambda_{eff}$. The vortex therefore exists only in an AdS$_3$ or Minkowski background. 
      
When $\Lambda_{eff}=0$, asymptotically we have $A(r)=D$ where $D$ here is positive (since a non-singular profile requires that $A(r)>0$). We also have $v_{eff}=v$. The differential equations governing the perturbations $\epsilon$ and $\beta$ are then
\begin{align}
&e^2 \,r \,v^2 \,\epsilon(r)+ D \left(\epsilon'(r)-r \epsilon''(r)\right)=0\\\nonumber\\
& 2\, r \,v^2 \lambda  \left(\alpha +v^2 \xi \right) \beta(r)-D \left(\alpha +v^2 \xi  (1+8 \xi )\right) \left(\beta'(r)+r \beta''(r)\right)=0
\end{align}
with solutions
\begin{align}
&\epsilon(r)=b \,e^{\frac{-e\,v\,r}{\sqrt{D}}}\,\sqrt{r}\label{eps2}\\
&\beta(r)= c\,e^{-v\,r\,\big(\frac{2 \,\lambda\,\alpha_{eff}}{D\,(\alpha_{eff} + 8\,v^2\,\xi^2)}\big)^{1/2}}\dfrac{1}{\sqrt{r}}
\label{beta2}
\end{align}
where $b$ and $c$ are positive constants. The above result is for a Minkowski background ($\Lambda_{eff}=0$) but where Einstein gravity and a non-minimal coupling term acts on the vortex. The exponential fall-off expressions \reff{eps2} and \reff{beta2} are similar to those found in fixed Minkowski spacetime \cite{Weinberg} and we recover them if we set $\xi=0$ and $D=1$.

\subsection{Plot of vortex profiles and magnetic field in AdS$_3$ for different $\xi$}

The parameters that appear in the Lagrangian density \reff{LDensity2} for the vortex are $\lambda$, $e$, $n$, $v$, $\alpha$, $\Lambda$ and $\xi$. The goal here is to determine how the vortex changes with the coupling $\xi$ and to observe what happens as we approach the critical couling $\xi_c$. How the vortex changes with the other parameters such as $\Lambda$, $n$ and $v$ has been studied elsewhere \cite{Ariel}. We therefore run numerical simulations for different values of $\xi$ with the other parameters held fixed; we set $\lambda=1$, $e=3$, $n=1$, $v=1$, $\alpha=1$, and $\Lambda=-1$. We work in natural units where $\hbar=c=1$. Though our parameters and quantities such as the radius, mass and magnetic field are quoted as numbers, they should be thought of as having a unit attached to them (except for the winding number $n$ which is a pure number)\footnote{In AdS$_3$ the appropriate length scale is the AdS length $\ell$. From \reff{A0}, the quantity $-\Lambda_{eff}\,r^2$ must be dimensionless. We quote $\Lambda_{eff}$ as a pure negative number but one should think of a unit $y$ attached to it so that $\Lambda_{eff} \times y =-1/\ell^2$. Therefore the unit attached to the radius $r$ is $y^{-1/2}$ which in terms of the AdS length is $(-\Lambda_{eff})^{1/2}\,\ell$. Note that the equation for $\epsilon(r)$ in \reff{beta} implies that $e\,v_{eff}/(-\Lambda_{eff})^{1/2}$is dimensionless. The quantity $\lambda/e^2$ is also dimensionless. The mass is proportional to $\alpha_{eff}= \alpha + v_{eff}^2 \,\xi$ and therefore the mass is expressed in units of the VEV squared which is $y^{1/2}$ and this can be expressed in terms of the inverse of the AdS length. The magnetic field is given by $B_m= \frac{\sqrt{A}\, a'}{e \,r}$ and since $A(r)$ and $a(r)$ are dimensionless, it has units of $y^{3/4}$which can be expressed in terms of the inverse of the AdS length to the power of $3/2$.} As we have seen, a negative $\Lambda$ automatically ensures that the asymptotic cosmological constant $\Lambda_{eff}$ will be negative. Our solutions in this section will therefore correspond to an AdS$_3$ background. Note that though $v$ and $\Lambda$ are held fixed, the VEV $v_{eff}$ and the cosmological constant $\Lambda_{eff}$ will change with $\xi$.

Recall that a critical coupling $\xi_c$ exists only if $v^4 \lambda +8 \alpha \Lambda$ is negative. With the above values for the parameters this latter quantity is negative (equal to $-7$) and therefore a critical coupling exists. It is given by \reff{Critical} and substituting the values of our parameters is equal to $\xi_c=2/21\approx 0.0952$ (the same value that appears in our plot of the VEV vs. $\xi$ in fig. \ref{VEV}). This implies that for $\xi\ge 2/21$ the VEV is zero and there is no vortex. We therefore obtained vortices for $\xi<2/21$.

\begin{figure}[!htb]
	\centering
		\includegraphics[scale=0.75]{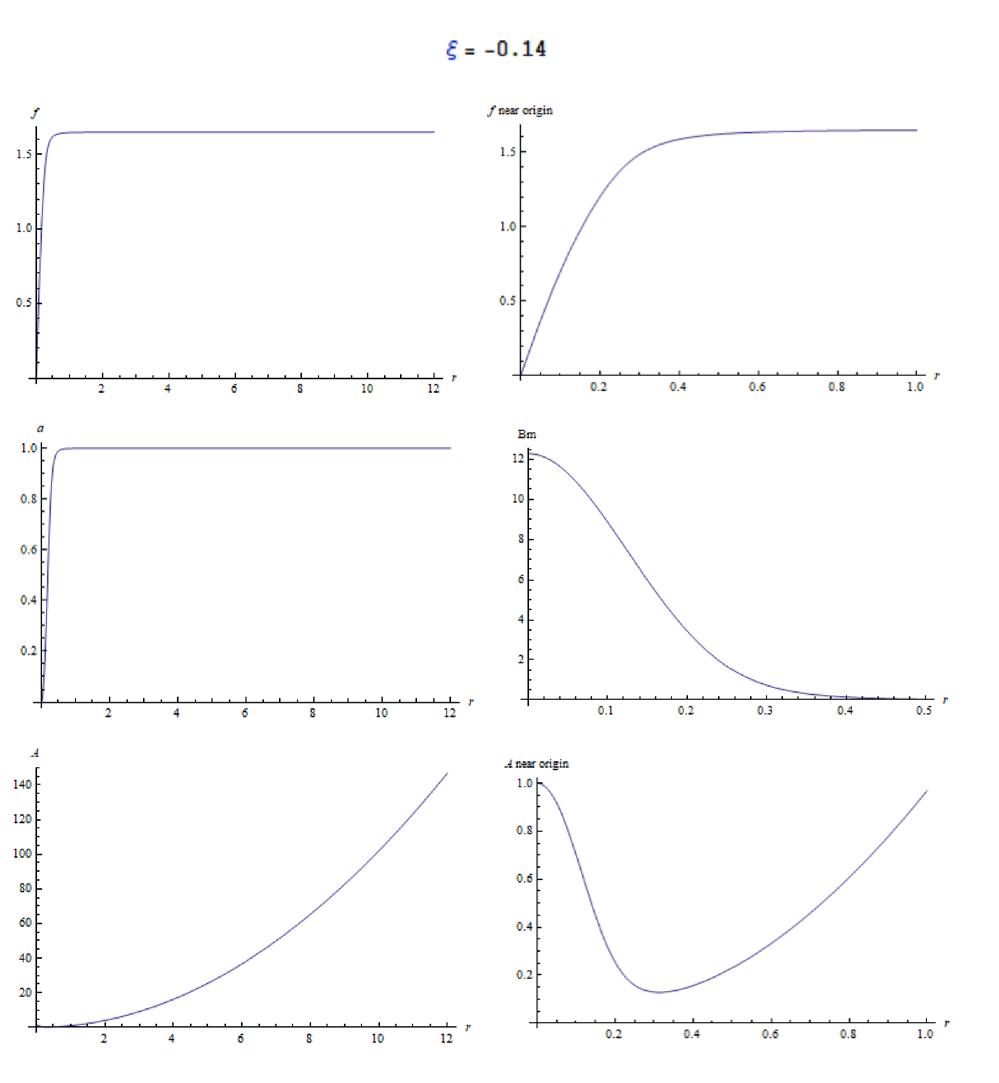}
		\caption{Case $\xi=-0.14$. This is the case with the lowest value of $\xi$ and the highest VEV (value where $f$ plateaus). The gauge field plateaus at $n=1$ which is the same value for all subsequent cases. The dip in the metric function $A(r)$ near the origin is the most pronounced of our sample. The magnetic field $B_m$ peaks at the origin and has the highest peak in our sample. The magnetic field also falls off the fastest (extends out the least). The plot of $f$ near the origin shows that $f$ plateaus quickly (does not extend much before reaching its VEV).}     
	\label{Graph1}
\end{figure}

\begin{figure}[!htb]
	\centering
		\includegraphics[scale=0.9]{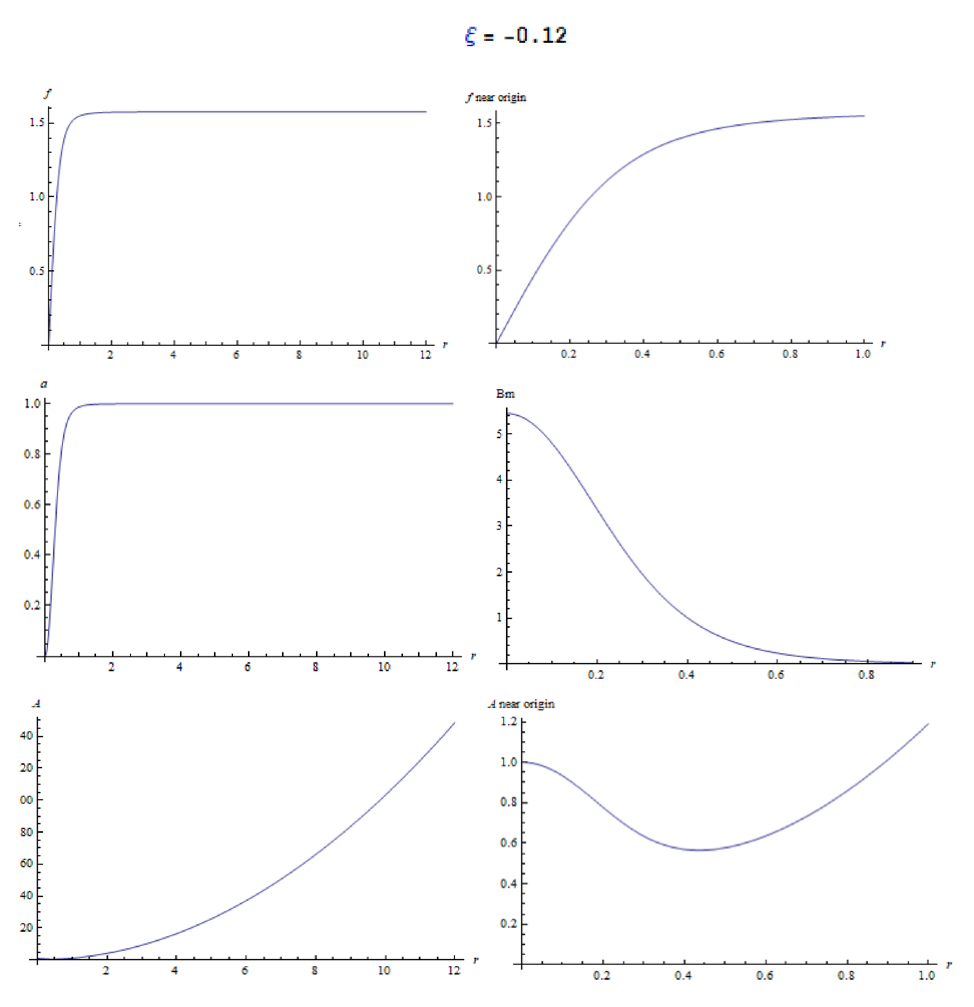}
		\caption{Case $\xi=-0.12$. This is the case with the next lowest value of $\xi$. $f$ plateaus at a lower VEV than $\xi=-0.14$ but it has the highest (ADM) mass in our sample. The dip in the metric function $A(r)$ near the origin is not as pronounced as in $\xi=-0.14$. The magnetic field $B_m$ at the origin is lower than for $\xi=-0.14$ but it falls off slower so that the magnetic flux turns out to be the same. Again, the plot of $f$ near the origin shows that $f$ plateaus quickly and hence has a small extension. }     
	\label{Graph2}
\end{figure}

\begin{figure}[!htb]
	\centering
		\includegraphics[scale=0.9]{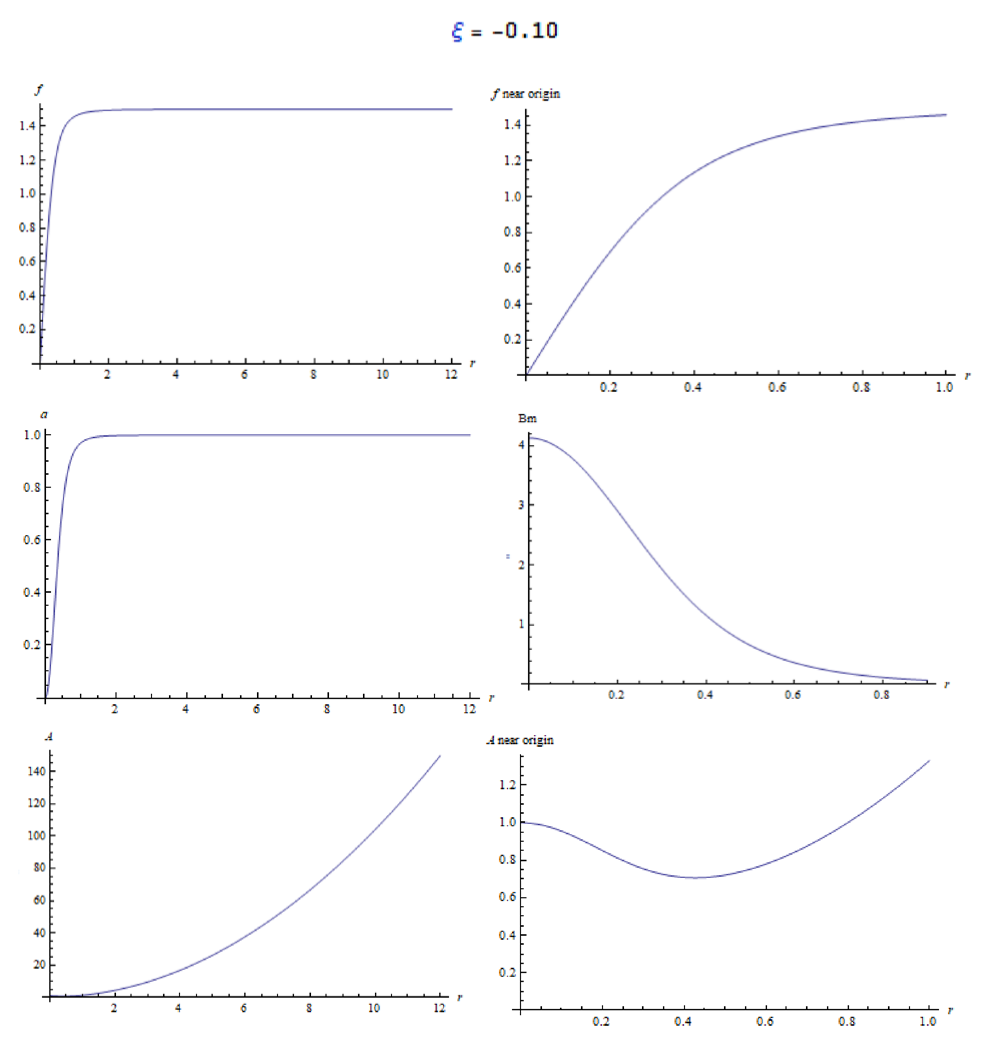}
		\caption{Case $\xi=-0.10$. $f$ plateaus at a lower VEV than the previous cases. The dip in the metric function $A(r)$ near the origin is still pronounced but not as much as in $\xi=-0.12$ or $\xi=-0.14$. The magnetic field $B_m$ at the origin is lower than for $\xi=-0.12$ but it falls off slower which yields the same magnetic flux as previous cases. The plot of $f$ near the origin shows that $f$ still plateaus relatively quickly though less fast than in previous cases. }     
	\label{Graph3}
\end{figure}

\begin{figure}[!htb]
	\centering
		\includegraphics[scale=0.9]{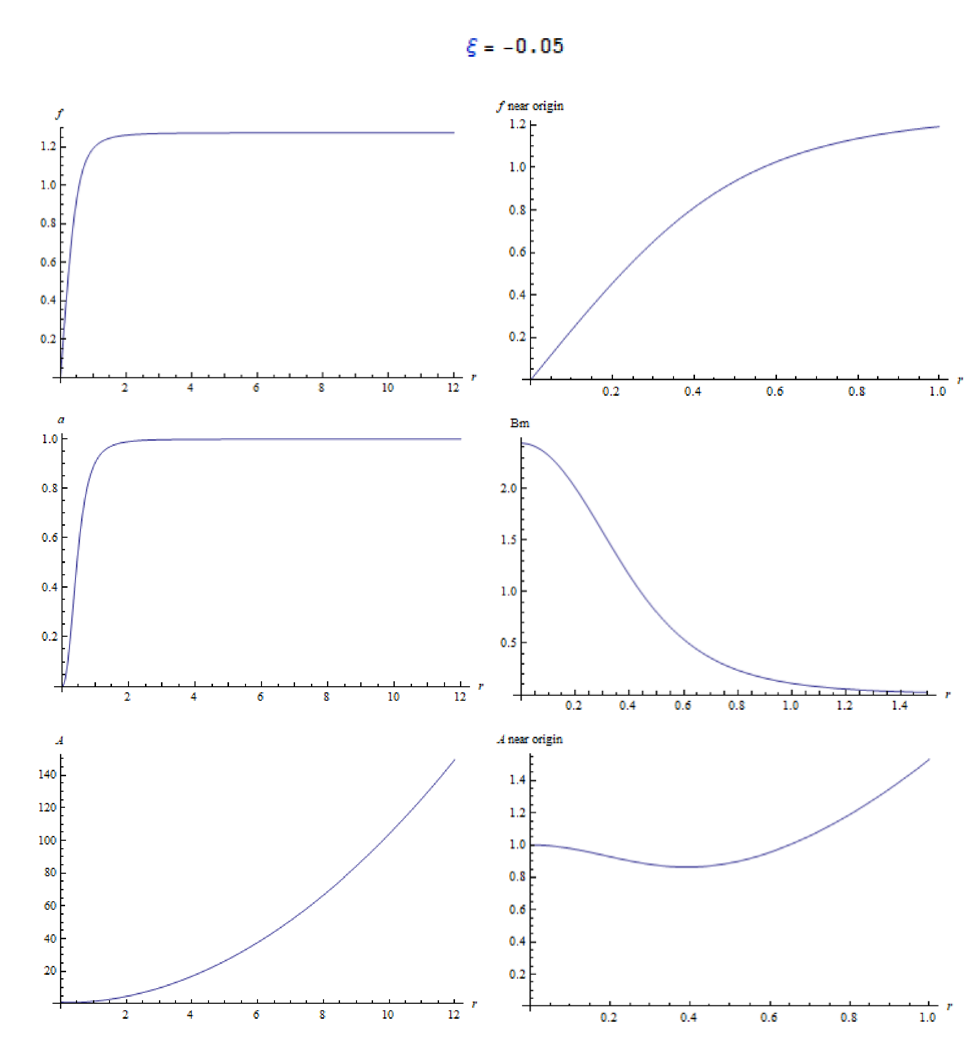}
		\caption{Case $\xi=-0.05$. $f$ plateaus at a lower VEV than the previous cases. The dip in the metric function $A(r)$ near the origin is visible but not as pronounced as in previous cases. The magnetic field $B_m$ has a profile that yields the same magnetic flux as previous cases. The plot of $f$ near the origin now shows that $f$ is no longer plateauing quickly (it needs to extend more before reaching its VEV).}     
	\label{Graph4}
\end{figure}

\begin{figure}[!htb]
	\centering
		\includegraphics[scale=0.9]{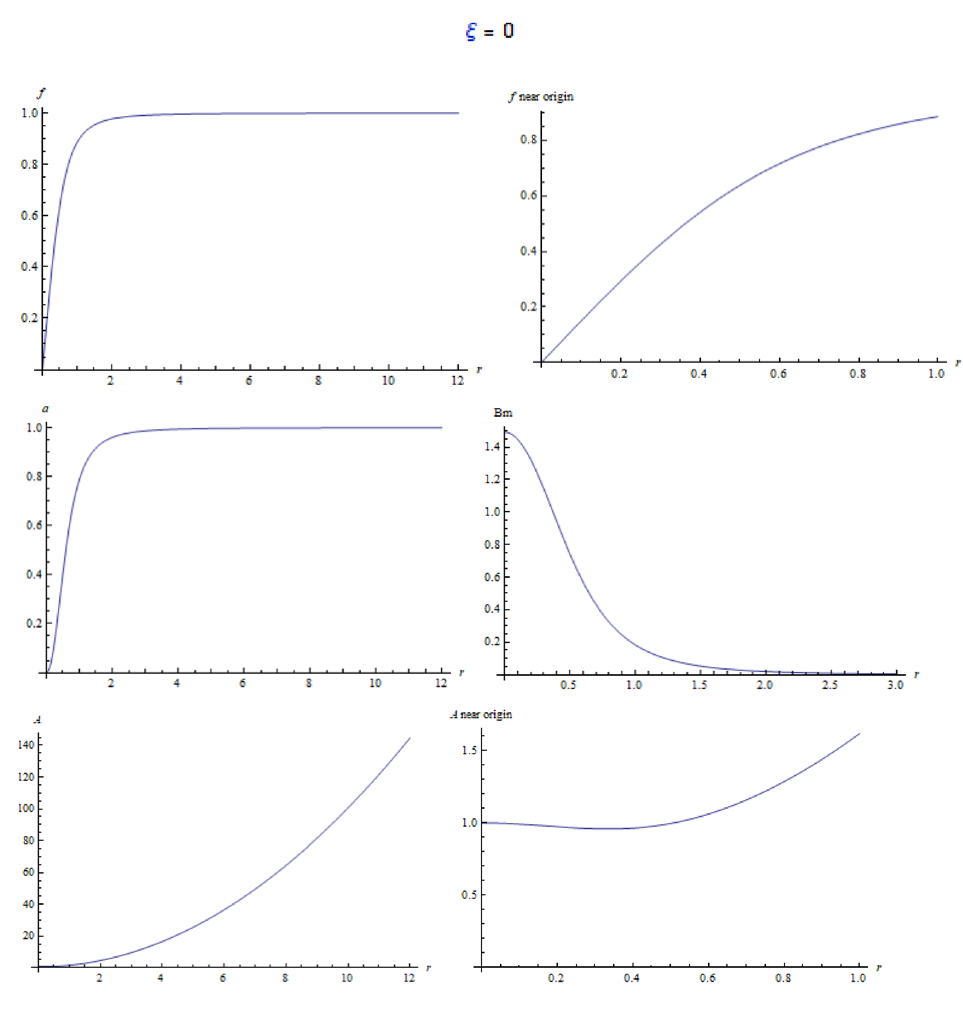}
		\caption{Case $\xi=0$. The non-minimal coupling term is turned off. The VEV is therefore equal to $v=1$. The dip in the metric function $A(r)$ near the origin is still visible. The magnetic field $B_m$ extends further out but yields the same magnetic flux as previous cases. The plot of $f$ near the origin shows that $f$ is still rising and requires more radial distance before it plateaus to its VEV).}     
	\label{Graph5}
\end{figure}

\begin{figure}[!htb]
	\centering
		\includegraphics[scale=0.8]{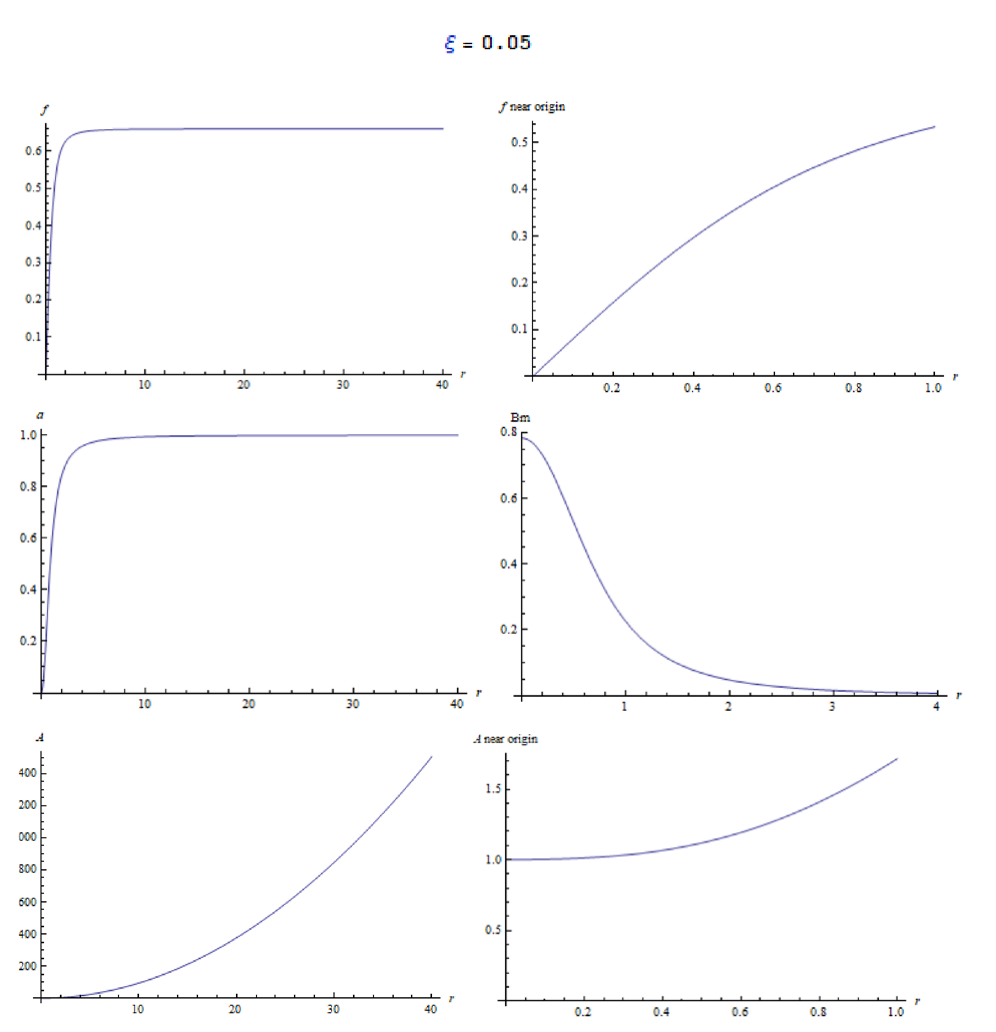}
		\caption{Case $\xi=0.05$. The regular plot of $f$ vs. $r$ now has a computational boundary of $R=40$ whereas in all previous cases it was $R=12$. This is because $f$ reaches its VEV now much slower and one needs to extend the computational boundary so that $f$ can reach its VEV to the same level of accuracy. The plot of $f$ near the origin shows that $f$ has a large slope and is also far from its plateau value so that it requires significantly more radial distance before it plateaus to its VEV. The numerical values of the metric function $A$ show that there is an extremely tiny dip right near the origin but this is not visible on the plot. The magnetic field $B_m$, just like $f$, extends further out than all previous cases.}     
	\label{Graph6}
\end{figure}

\begin{figure}[!htb]
	\centering
		\includegraphics[scale=0.9]{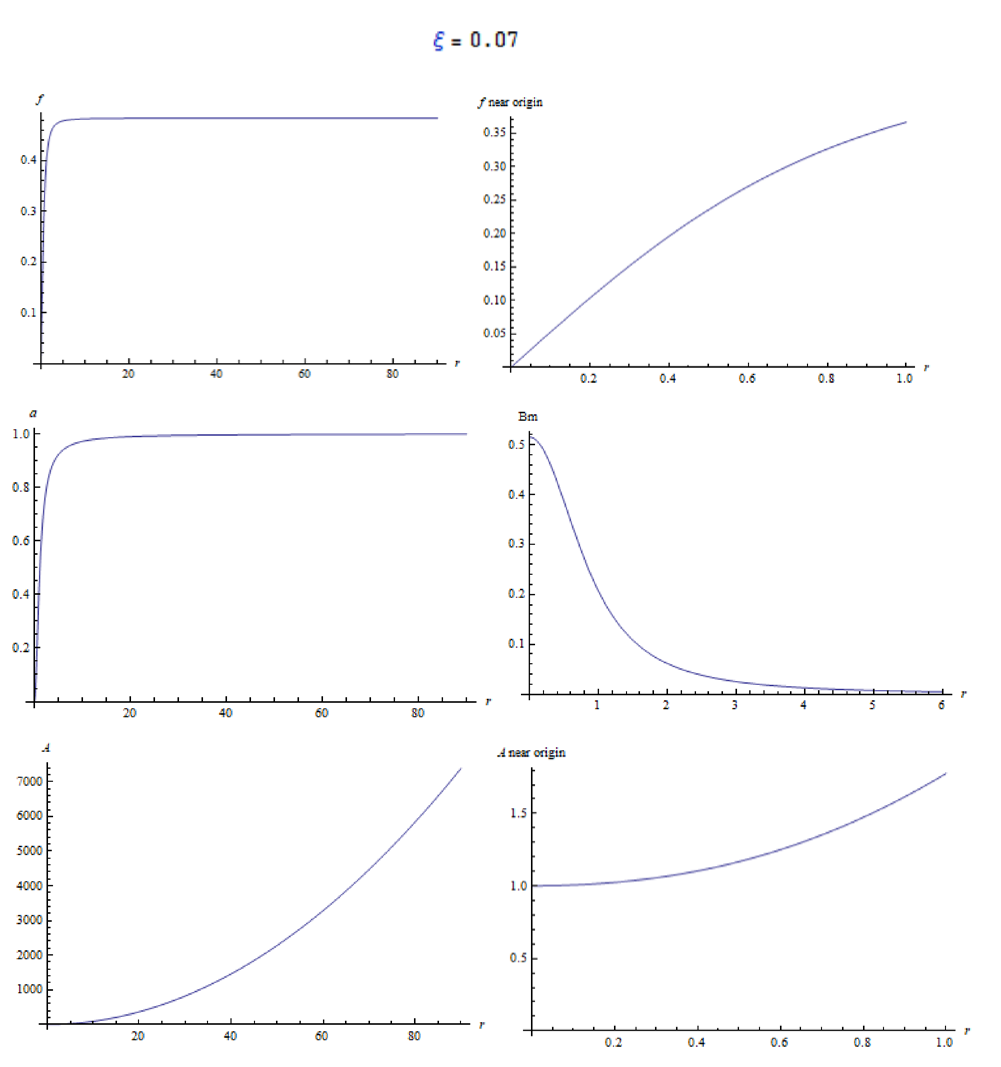}
		\caption{Case $\xi=0.07$. The regular plot of $f$ vs. $r$ has a computational boundary of $R=90$. The plot of $f$ near the origin shows that $f$ is now quite far from its plateau value. It requires now a larger radial distance before it plateaus to its VEV.  There is no longer any dip in the metric function $A$: the numerical values show $A(r)$ increases monotonically. The magnetic field $B_m$, just like $f$, extends out much further than previously.}     
	\label{Graph7}
\end{figure}

\begin{figure}[!htb]
	\centering
		\includegraphics[scale=0.75]{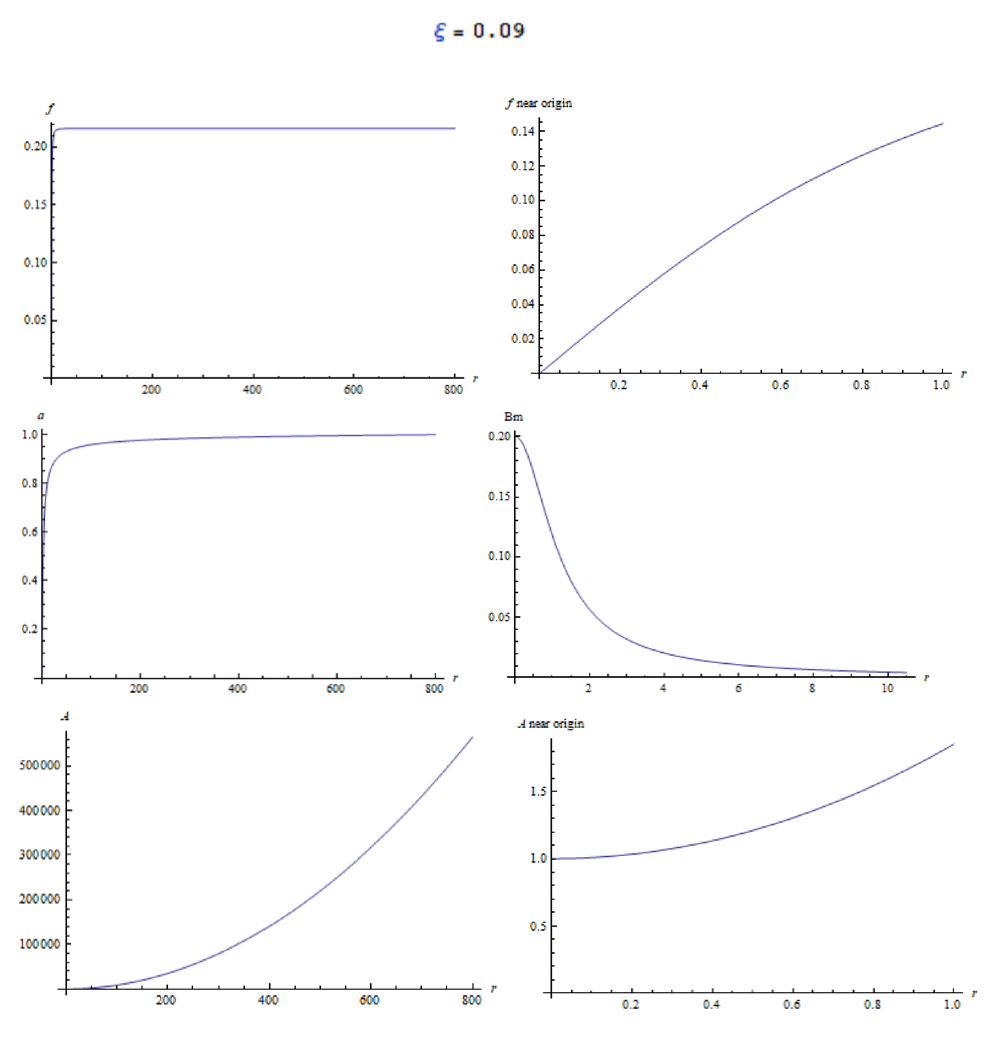}
		\caption{Case $\xi=0.09$. This is the second largest $\xi$ in our sample and we are now getting quite close to the critical coupling $\xi_c\approx 0.0952$ where the derivative of the VEV with respect to $\xi$ diverges. The change from $\xi=0.07$ to $\xi=0.09$ is therefore large. The regular plot of $f$ vs. $r$ has a significantly larger computational boundary of $R=800$. The plot of $f$ near the origin shows that $f$ is very far from its plateau value. It requires now a very large radial distance before it plateaus to its VEV.  Again, there is no longer any dip in the metric function $A$ and it increases monotonically. The magnetic field $B_m$, just like $f$, extends out again much further than previously.}     
	\label{Graph8}
\end{figure}

\begin{figure}[!htb]
	\centering
		\includegraphics[scale=0.75]{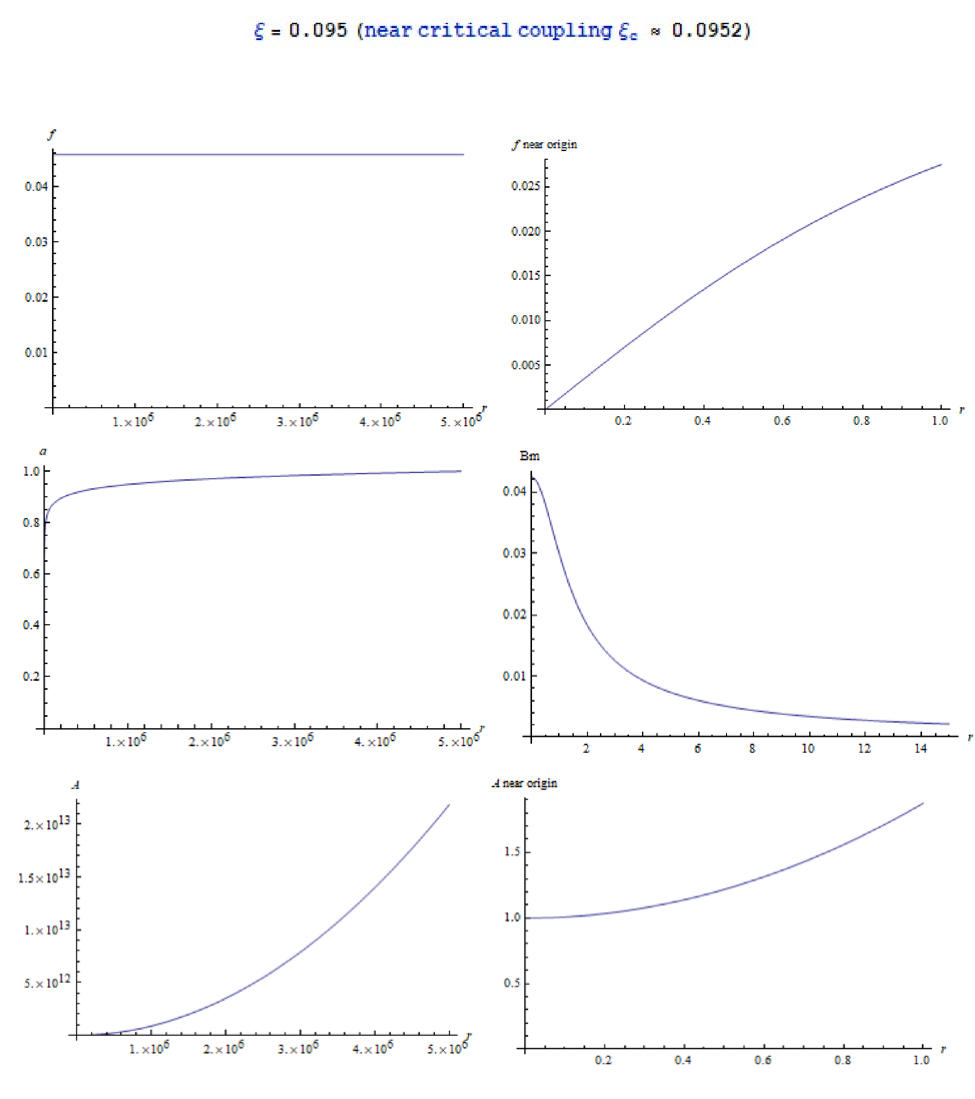}
		\caption{Case $\xi=0.095$. This is the largest $\xi$ in our sample and is very close to the critical coupling $\xi_c\approx 0.0952$. Since we are near the critical point, the change from $\xi=0.09$ to $\xi=0.095$ is very large. The plot of $f$ near the origin shows that $f$ is again very far from its plateau value; this is why the regular plot of $f$ vs. $r$ requires an extremely large computational boundary of $R=5 \times 10^6$. This is the radius required for $f$ to reach its VEV to the same level of accuracy as the other cases. The ``extension" of $f$ (a measure of the radius required to reach the VEV) has therefore increased enormously as $\xi$ approached near the critical coupling $\xi_c$ and this is analogous to the divergence of the coherence length in GL mean-field theory as one approaches the critical temperature $T_c$.}     
	\label{Graph9}
\end{figure}

We considered nine values of the coupling $\xi$ that ranged from $-0.14$ to $0.095$ (close to the upper bound $\xi_c$) which includes the case $\xi=0$. We present below figures \ref{Graph1}-\ref{Graph9}, one for each value of the couplings in order of increasing $\xi$. Each figure contains plots of the scalar field $f(r)$, the gauge field $a(r)$, the metric function $A(r)$ and the magnetic field $B_m(r)$. We also made separate plots of $f$ and $A$ that focus on the core region near the origin where the fields undergo significant change. There are therefore six plots associated with each value of $\xi$. We also present some numerical results in table format. In table \ref{Table} we present the following data for each value of $\xi$: the theoretically expected and numerically obtained value of the VEV $v_{eff}$ and cosmological constant $\Lambda_{eff}$, the (ADM) mass of the vortex, the peak value of the magnetic field at the origin and the numerically integrated magnetic flux. 

In table \ref{Table}, the formula \reff{Veff} for the VEV $v_{eff}$ matched almost exactly (to within three and four decimal places) the value where $f$ plateaued numerically. Similarly, our formula \reff{Leff} for the cosmological constant $\Lambda_{eff}$ matched almost exactly (again to within three and four decimal places) the numerical value of the asymptotic cosmological constant. This provides strong confirmation of both our analytical formulas and numerical simulation. In figures \ref{Graph1}-\ref{Graph9}, the magnetic field $B_m$ always peaks at the origin and then falls off with radius towards zero. As $\xi$ increases and approaches closer to the critical coupling, the value of the peak magnetic field decreases (see plot in fig. \ref{BPlot}) but the magnetic field extends further out since it falls off to zero more slowly. As a consequence, the magnetic flux obtained numerically by integrating over the magnetic field profile remained constant as $\xi$ changed (see table \ref{Table}) and matched exactly (to within three or four decimal places) the expected theoretical value of $\Phi=2\,\pi\,n/e=2 \pi/3=2.0944$ (where we substituted $n=1$ and $e=3$). That this numerically integrated magnetic flux remained constant across different magnetic field profiles provides another strong check on our numerical simulation.   

\begin{figure}[!htb]
		\includegraphics[scale=0.3]{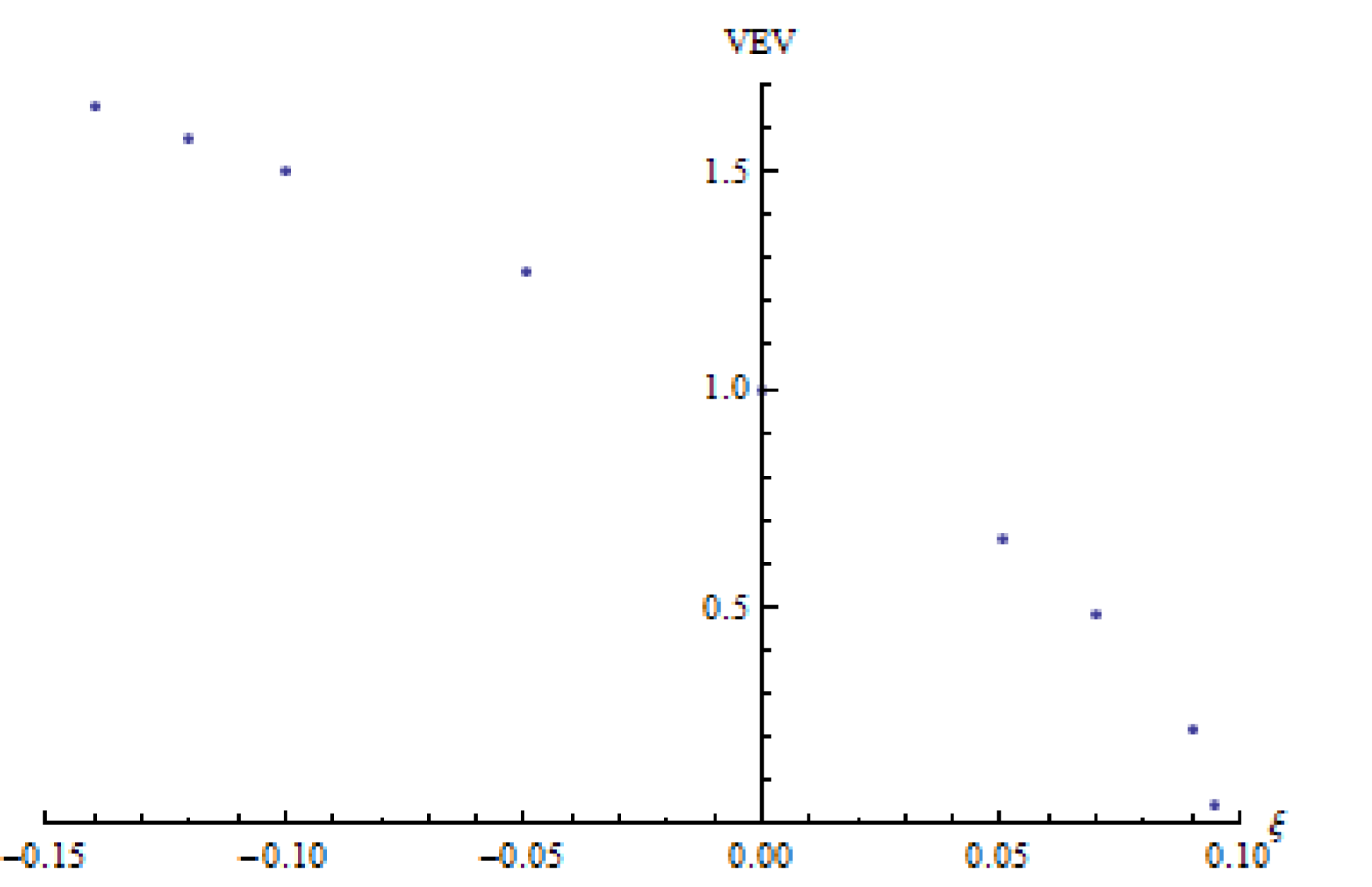}
		\caption{Plot of the numerical value of the VEV vs. $\xi$. The data points trace out a curve similar to the plot in fig. \ref{VEV} of the VEV vs. $\xi$ that was obtained theoretically and similar to the curve in fig. \ref{Order} of the order parameter vs. temperature in GL mean-field theory. The VEV decreases monotonically and its slope gets steeper (more negative) as $\xi$ increases towards the critical coupling. The data points near $\xi_c$  obey the power law $v_{eff}\propto (\xi_c-\xi)^{1/2}$ (see body of text above for exact comparison); this confirms that our system undergoes critical phenomena with a critical exponent of $1/2$.}     
	\label{VEVPlot}
\end{figure}

\begin{figure}[!htb]
		\includegraphics[scale=0.3]{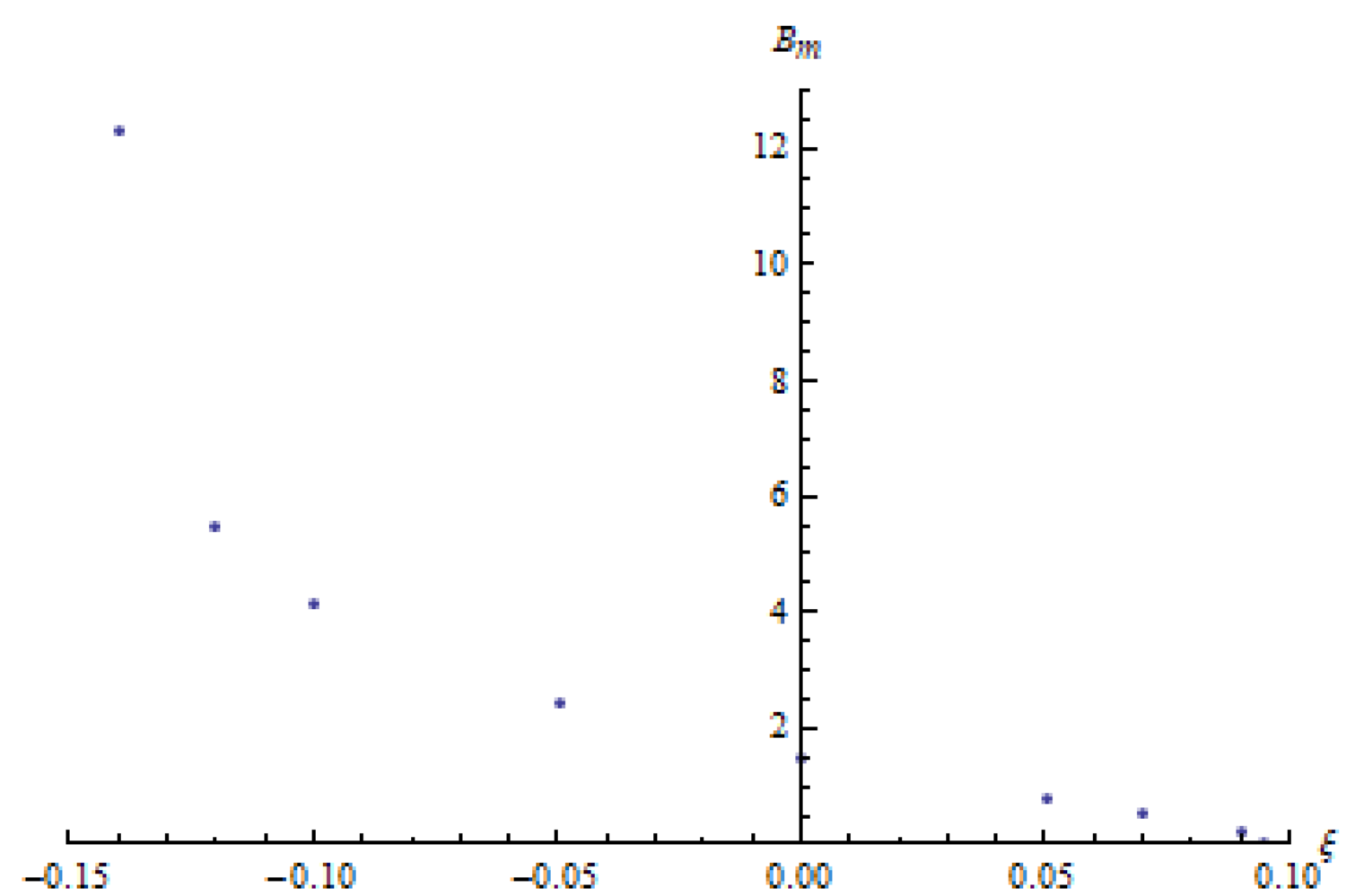}
		\caption{Plot of peak magnetic field vs. $\xi$. Like the VEV, it decreases monotonically as $\xi$ increases but in sharp contrast to the VEV, its slope gets flatter (less negative) as $\xi$ increases towards the critical coupling. Therefore, the peak magnetic field does not act like an order parameter.}     
	\label{BPlot}
\end{figure}

In table \ref{Table}, the VEV monotonically decreases from a value of $1.6475$ at $\xi=-0.14$ to a value of $0.04584$ at $\xi=0.095$. We plot the nine data points in fig. \ref{VEVPlot} and they trace out a curve similar to the plot in fig. \ref{VEV} of the VEV vs. $\xi$ obtained theoretically and hence also similar to the plot in fig. \ref{Order} of the order parameter vs. temperature in GL mean-field theory. We now verify numerically that the data points in our sample that are close to the critical coupling $\xi_c=2/21$ follow the power law with critical exponent $1/2$ that we previously derived for $\xi$ near $\xi_c$ i.e. $v_{eff}=k\,(\xi_c-\xi)^{1/2}$ where $k=\tfrac{v}{\sqrt{\xi_c + (2\,v^2\,\xi_c^2)/\alpha}}$  (see \reff{expansion}). For the values of our parameters we obtain $k=2.96985$. For $\xi=0.095$, which is the closest data point to $\xi_c$ in our sample, we obtain $k\,(\xi_c-\xi)^{1/2}=0.04583$ which matches almost exactly our numerical result of $0.04584$ for the VEV quoted in table \ref{Table}. Another data point we can consider is $\xi=0.09$ as it is not that far off from the critical coupling. This yields $k\,(\xi_c-\xi)^{1/2}=0.2149$ which still matches quite closely our numerical result of $0.2161$. This constitutes a quantitative confirmation that the non-minimally coupled vortex in AdS$_3$ undergoes critical phenomena with exponent $1/2$ at the critical coupling $\xi_c$.

We mentioned above that the magnetic field extends further out as $\xi$ increases towards the critical coupling $\xi_c$. The same thing happens with the scalar field $f$. For cases $\xi=-0.14$, $\xi=-0.12$ and $\xi=-0.10$, $f$ can be seen to roughly plateau already ``near the origin" (see plots of $f$ ``near the origin" in figures \ref{Graph1}-\ref{Graph3}). At higher $\xi$, $f$ has not plateaued yet near the origin (see plots of $f$ ``near the origin" in figures \ref{Graph4}-\ref{Graph9}). This implies that it must extend further out to reach its VEV. In particular, as $\xi$ approached near the critical coupling $\xi_c$, the regular plot of $f$ vs. $r$ has to be extended to drastically larger radii to accommodate the fact that $f$ plateaus so much more slowly. We will discuss the extension of the scalar field (and of the magnetic field) in more detail in the next subsection.      

If the local matter density in the core region of the vortex is high enough it causes the metric function $A(r)$ near the origin to have a noticeable dip: the metric starts at $A=1$ at the origin $r=0$, dips below unity in the core region, reaches a minimum that is above zero before increasing to reach its asymptotic $r^2$ dependence. The dip can be seen in the plot of $A$ ``near the origin" and the asymptotic $r^2$ dependence is more evident in the regular plot of $A$ vs. $r$. The plots of the metric function $A(r)$ near the origin in figures \ref{Graph1} to \ref{Graph9} reveals that the dip monotonically decreases as $\xi$ increases and is most pronounced at $\xi=-0.14$. This implies that the local matter density in the core region is greatest for $\xi=-0.14$. Though $A$ in this case dips the closest to zero (i.e. its minimum is smallest) it does not cross zero. If $A$ crossed zero, this would signal black hole formation and a singularity. However, our non-singular initial conditions prevents one from constructing vortices beyond a local matter density where gravity becomes so strong that the scalar field is no longer able to reach its asymptotic plateau value. The fact that $f$ is fixed to be zero at the origin prevents one from constructing vortex solutions when gravity's effect gets too strong. This places a lower bound on $\xi$; for the values of our parameters, we were not able to construct non-singular vortices roughly below $\xi=-0.14$. This lower bound was reached way before the lower bound set by the condition $\alpha_{eff}=\alpha +v_{eff}^2 \xi>0$. With $v_{eff}$ given by \reff{Veff} and using the values of our parameters, one can readily check that this would have occurred at the much lower value of $\xi=-0.26$. 

In table \ref{Table} one can see that the ADM mass is highest at $\xi=-0.12$ and decreases afterwards as $\xi$ increases towards $\xi=0.095$. There is one case that does not follow this trend in masses. The ADM mass at $\xi=-0.14$ is actually lower than the mass at $\xi=-0.12$ (the data points of mass vs. $\xi$ is plotted in fig. {MPlot} and the curve illustrates nicely the trend in masses). The case $\xi=-0.14$ has the highest VEV which would seem to imply that it should have the highest mass (vortices with higher VEV will usually have more mass in fixed Minkowski spacetime \cite{Weinberg}). Why then is the mass lower for $\xi=-0.14$ than for $\xi=-0.12$? This is due to the fact that the ADM mass receives contributions not only from matter but also from the negative binding energy of the gravitational field (see section 3.9 on ``Thin-shell collapse" in \cite{Poisson} for a clear illustration of this). The metric field $A(r)$ near the origin for $\xi=-0.14$ (fig. \ref{Graph1}) has a more pronounced dip than for $\xi=-0.12$ (fig. \ref{Graph2}). So the negative gravitational binding energy is significant enough in $\xi=-0.14$ to yield a lower ADM mass than in $\xi=-0.12$.
\clearpage
          
\clearpage
\pagebreak



\begin{table}[b]
	\centering
		\includegraphics[scale=0.35]{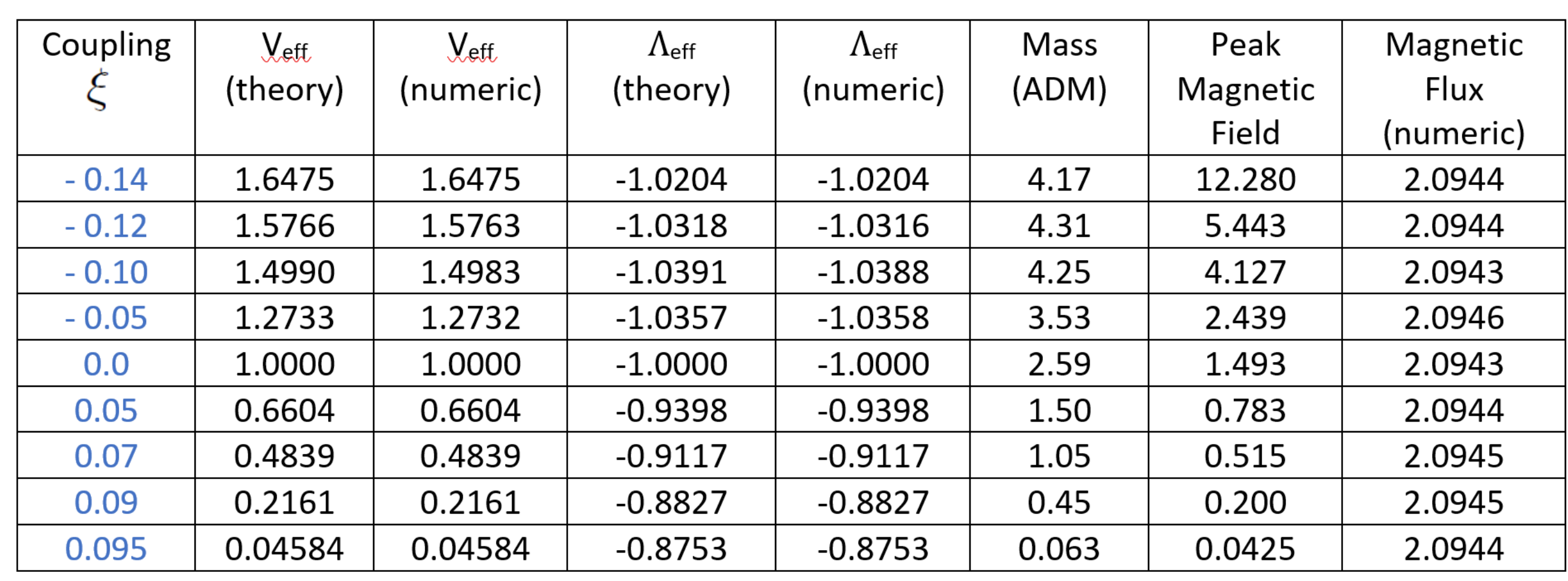}
		\caption{We present data for $\xi$ ranging from $-0.14$ to $0.095$ (near the critical coupling $\xi_c=2/21\approx 0.0952$). The theoretically predicted values of the VEV $v_{eff}$ and cosmological constant $\Lambda_{eff}$ calculated using \reff{Veff} and \reff{Leff} respectively match the numerical values to within three or four decimal places. The peak value of the magnetic field occurs at the origin and also decreases monotonically as $\xi$ increases. The magnetic flux obtained by integrating numerically over the magnetic field profile remains constant despite the different profiles and its numerical value matches the theoretically expected value of $\Phi=2\,\pi\,n/e=2.0944$ to within three or four decimal places. This provides a very strong check on our numerical simulation. The ADM mass increases from $\xi=0.095$ to $\xi=-0.12$ but this trend does not extend all the way to $\xi=-0.14$; this is due to a significant negative gravitational binding energy in the case of $\xi=-0.14$ (see body of text for more details).}     
	\label{Table}
\end{table}

\begin{figure}[!htb]
		\includegraphics[scale=0.3]{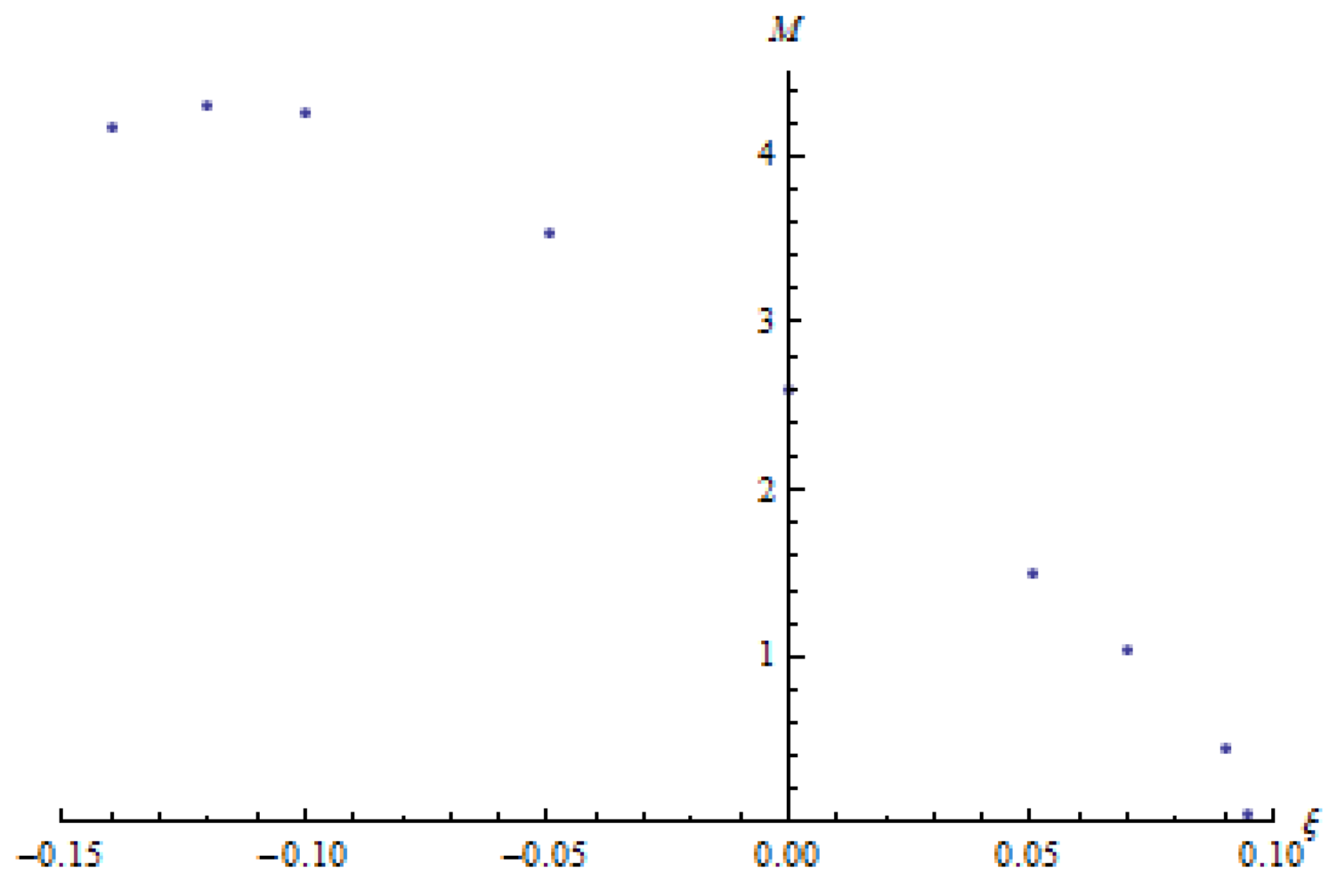}
		\caption{Plot of mass $M$ vs. $\xi$. Near the critical coupling, this plot looks similar to the one for the VEV; in particular, its slope gets steeper (more negative) as $\xi$ increases towards the critical coupling. The mass increases as $\xi$ decreases but unlike the VEV, this trend stops when we get to the most negative point, $\xi=-0.14$, where the mass is less than for $\xi=-0.12$ due to gravity's effect (see discussion in body of text).}     
	\label{MPlot}
\end{figure}
\FloatBarrier
\subsubsection{Extension of scalar field and magnetic field and divergence at critical coupling $\xi_c$} 

We have already mentioned that as $\xi$ increases towards the critical coupling, the scalar field and magnetic field extend further out. In the case of the scalar field, this means it rises slower and plateaus at its VEV over a longer radius. For the magnetic field, it means that starting from its peak at the origin, it decreases towards zero in a slower fashion, again over a longer radius. In short, the core region of the vortex occurs over a longer spartial range as $\xi$ gets larger.

\begin{figure}[!htb]
		\includegraphics[scale=0.3]{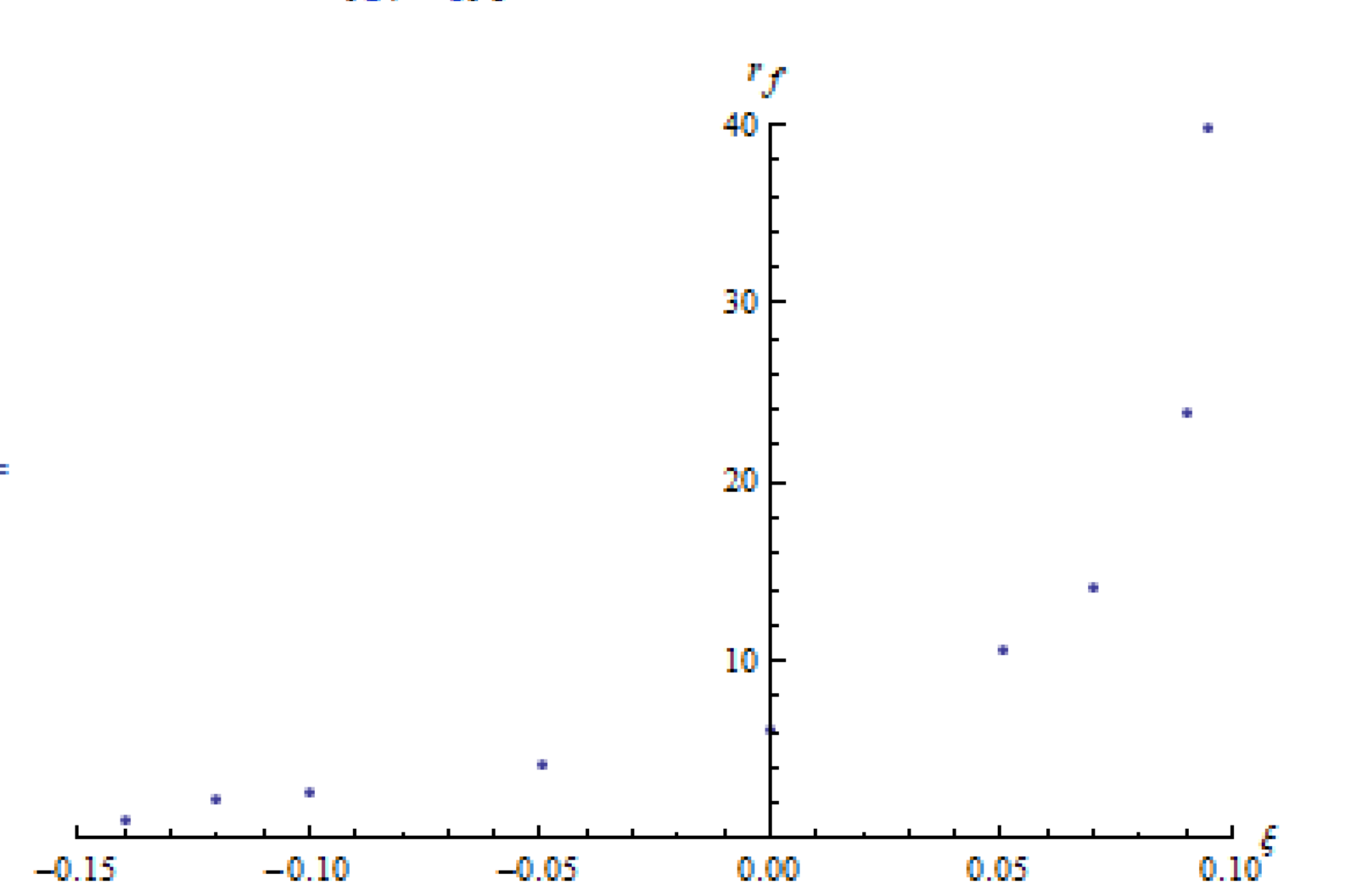}
		\caption{Extension of the scalar field $f(r)$ as a function of $\xi$. Note the rapid increase in the extension as one approaches near the critical coupling $\xi_c\approx 0.0952$. The extension is expected to diverge at the exact value of $\xi_c=2/21$.}     
	\label{Extf}
\end{figure}

\begin{figure}[!htb]
		\includegraphics[scale=0.3]{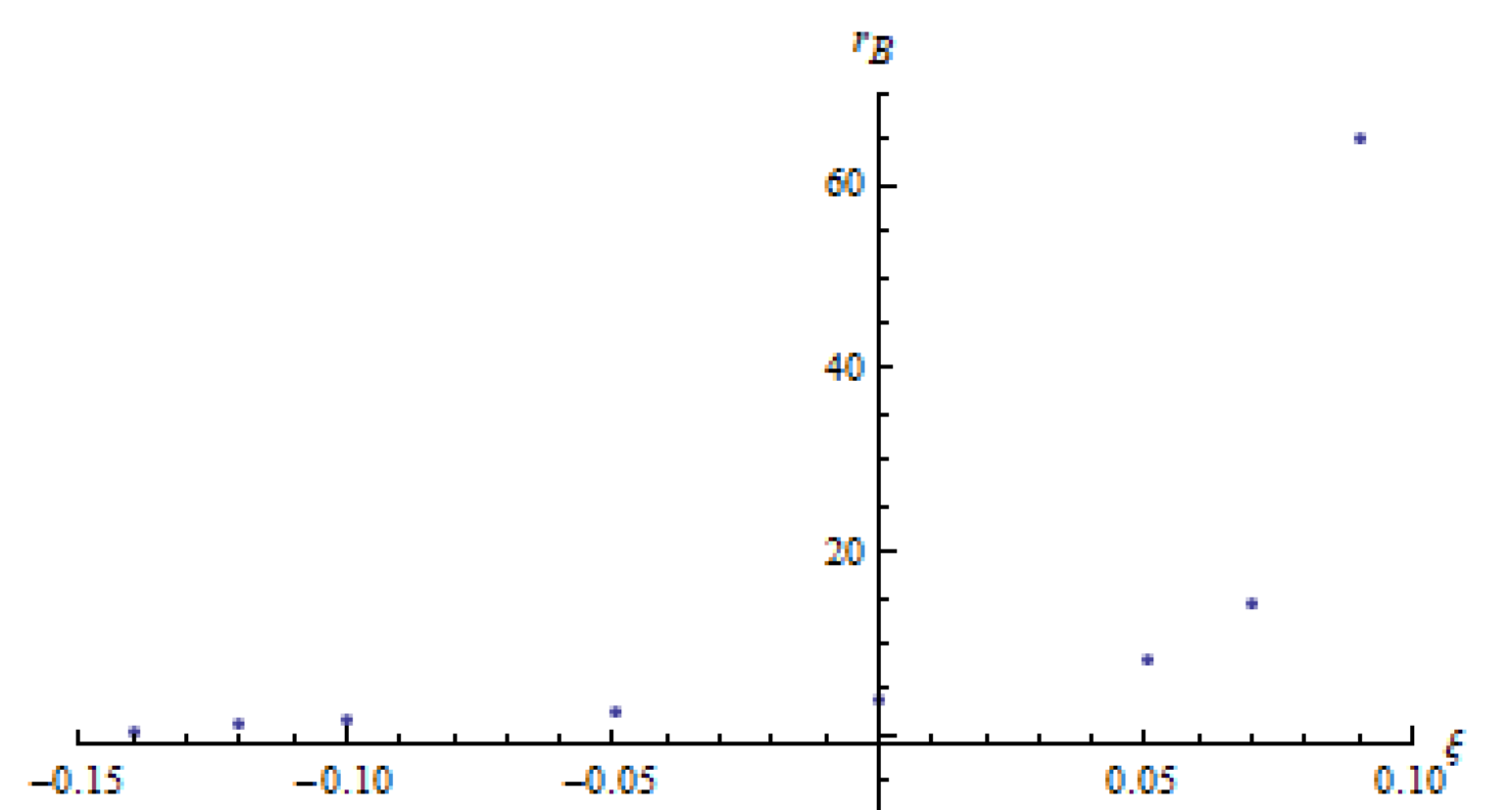}
		\caption{The extension of the magnetic field $B_m(r)$ as a function of $\xi$. Here also there is a rapid increase in the extension as one approaches near the critical coupling $\xi_c\approx 0.0952$. The extension is expected to diverge at the exact value of $\xi_c=2/21$.}     
	\label{ExtB}
\end{figure}

To make this more quantitative, we will define the extension $r_f$ of the scalar field to be the radius where it reaches $99.9\%$ of its VEV and define the extension $r_B$ of the magnetic field to be the radius where it has fallen to $0.1\%$ of it its peak value (i.e. decreased by $99.9\%$ from its peak at the origin). We plot in fig \ref{Extf} the scalar field extension $r_f$ vs. $\xi$ and in fig. \ref{ExtB} the magnetic field extension $r_B$ vs. $\xi$. In both cases, there is a very rapid increase in the extension when $\xi$ is near the critical coupling $\xi_c$. We will see that the extension actually diverges at the exact value of $\xi=\xi_c=2/21$. This is reminiscent of the divergence of the coherence length in GL mean-field theory at the critical temperature $T_c$. 

We will now show analytically that $f(r)$ approaches the VEV in the slowest fashion possible in the limit when $\xi$ approaches $\xi_c$. If we let $f(r)=v_{eff}-\beta(r)$ asymptotically, we know that $\beta(r)$ is given by \reff{beta} which we rewrite for convenience below 
\begin{align}
\beta(r)&= c\, r^{-1-\Big[\dfrac{-\alpha_{eff}\,\Lambda_{eff} + 2 \,\alpha_{eff}\,
v_{eff}^2 \,\lambda - 64 \,v_{eff}^2 \,\Lambda_{eff} \,\xi^2}{-\alpha_{eff}\,\Lambda_{eff} - 16 \,v_{eff}^2 \,\Lambda_{eff} \,\xi^2}\Big]^{1/2}}\nonumber\\
&= c\, r^{-1-P^{1/2}}
\label{beta3}
\end{align}
where $P$ is the quantity in square brackets. Since $\alpha_{eff}>0$ and $\Lambda_{eff}<0$, all the terms in the numerator and denominator in the square brackets are positive. It should be clear that $P \ge 1$. We have that $\beta$ approaches zero asymptotically as $1/ r^{1+P^{1/2}}$. When $\xi\to\xi_c$, we have that $v_{eff}\to 0$ and $P\to 1$. Therefore, as $\xi \to \xi_c$, $\beta$ decreases as $1/r^2$ asymptotically which is the slowest fall-off it can have which translates to the slowest approach that $f$ can have  towards its VEV.   

Now $r_f$ is the extension, defined as the radius where $f=0.999 \,v_{eff}$ so that $r_f^{1+P^{1/2}}$ is proportional to $1/(0.001\,v_{eff})$. This diverges as $\xi\to\xi_c$ since $v_{eff}\to 0$. It is therefore expected that the extension $r_f$ diverges at the critical coupling $\xi_c$ in accordance with the trend in fig. \ref{Extf}. 

Asymptotically we have that $a(r)=n-\epsilon(r)$ where $\epsilon$ is given by \reff{eps}. The magnetic field is given by $B_m=\sqrt{A(r)}\,a'(r)/(e\,r)$. Asymptotically, $A(r) \to -\Lambda_{eff}\,r^2$ and $a'(r) \to -\epsilon'(r)$ so that $B_m$ falls off asymptotically as $r^{-\frac{e\,v_{eff}}{(-\Lambda_{eff})^{1/2}}-1}$. As $\xi \to \xi_c$, we have that $v_{eff} \to 0$ so that $B_m$ falls off as $1/r$ which is the slowest fall-off possible. The extension $r_B$ is therefore proportional to the inverse of the peak magnetic field as $\xi \to \xi_c$. Our numerical results show that the peak value of the magnetic field at the origin keeps decreasing (towards zero) as $\xi\to \xi_c$ so that the extension $r_B$ tends to infinity. This agrees with the fact that the magnetic flux can remain constant as the peak magnetic field at the origin decreases to zero only if the magnetic field has an infinite extension.      

\subsection{Plot of vortex profiles and magnetic field in asymptotically Minkowski spacetime}   

We now consider the role the coupling $\xi$ plays for the case of asymptotically Minkowski spacetime. This corresponds to $\Lambda=0$ which as we have seen, implies $\Lambda_{eff}=0$. As previously mentioned, there is no critical coupling for asymptotically Minkowski spacetime. The VEV is expected to remain constant at $v_{eff}=v=1$ and the cosmological constant is expected to remain at $\Lambda_{eff}=0$ i.e. the VEV $v_{eff}$ and $\Lambda_{eff}$ have no dependence on $\xi$ in contrast to the AdS$_3$ case. We run numerical simulations for different values of $\xi$ with the same set of parameters as before: $\lambda=1$, $e=3$, $n=1$, $v=1$ and $\alpha=1$. The only difference is that $\Lambda=0$ now (instead of the $\Lambda=-1$ we used in the AdS$_3$ case). We work again in natural units. As before, the parameters and quantities like the radius, mass and magnetic field are quoted as numbers but one should think of a unit attached to them\footnote{In Minkowski spacetime, the appropriate length scale is set by the VEV $v$. In particular, $e\,v\,r$ is dimensionless where $r$ is the radius. Though $e$ and $v$ are quoted as numbers one should think of $e\,v$ as having a unit $x$ of dimension [L]$^{-1}$ attached to it. It follows then that the radius $r$ has units of $x^{-1}$ which has the correct dimensions of [L]. The mass is proportional to the VEV squared and is therefore expressed in units of $x$ which has the correct dimension of $[L]^{-1}$. The magnetic field $B_m= \frac{\sqrt{A}\, a'}{e \,r}$ is expressed in units of $x^{3/2}$ which has the correct dimensions of [L]$^{-3/2}$. As before, $\lambda/e^2$ is dimensionless.}.       

We made plots for five different cases: $\xi=\{-0.4,-0.2,0.0,0.2,0.4\}$. The plots of the scalar field $f$ and the gauge field $a$ all plateau at unity regardless of $\xi$. We also plot the magnetic field whose profile depends on $\xi$. The most important plot by far is the one for the metric $A$ which plateaus asymptotically to a constant value (which we previously labeled $D$). The profile of the metric here (starting at unity at the origin and then plateauing to $0<D<1$) is in stark contrast to the AdS$_3$ case where the metric had an $r^2$ dependence asymptotically. The constant $D$ can only be obtained numerically (by running the simulation) and it changes with $\xi$. Since the deficit angle depends on $D$ via \reff{delta}, the deficit angle depends on $\xi$. We also calculate the mass $M_{flat}$ of the vortex via \reff{M2}. The constant $D$, the deficit angle $\delta$, the mass $M_{flat}$ as well as the peak value of the magnetic field are presented in table \ref{Table2}. In $2+1$-dimensional General Relativity in asymptotically Minkowski spacetime, there is the classic result due to Deser et al. \cite{Deser} that a point mass produces a deficit angle proportional to the mass. The ratio of mass to deficit angle is equal to $2\ \alpha=1/(8\,\pi\,G)$ and is a constant since Newton's constant $G$ does not change as the mass changes. In contrast, for the vortex with non-minimal coupling, the ratio of mass to deficit angle is not constant but depends on $\xi$: it is equal to $2\alpha_{eff}=2(\alpha+v^2\,\xi)=2\,(1+\xi)$ where we substituted the values $\alpha=1$ and $v=1$ for our parameters. A striking consequence is that it is possible for a larger mass to actually produce a smaller deficit angle compared to a smaller mass. For example, in table \ref{Table2}, the case at $\xi=-0.4$ has the largest deficit angle of 1.894 rad in our sample and has a mass of 2.273 whereas the case at $\xi=0.4$ has a smaller deficit angle of 1.225 rad but the largest mass of 
3.430 in our sample which is roughly 1.5 times greater than our former case. 
\begin{figure}[!htb]
		\includegraphics[scale=0.7]{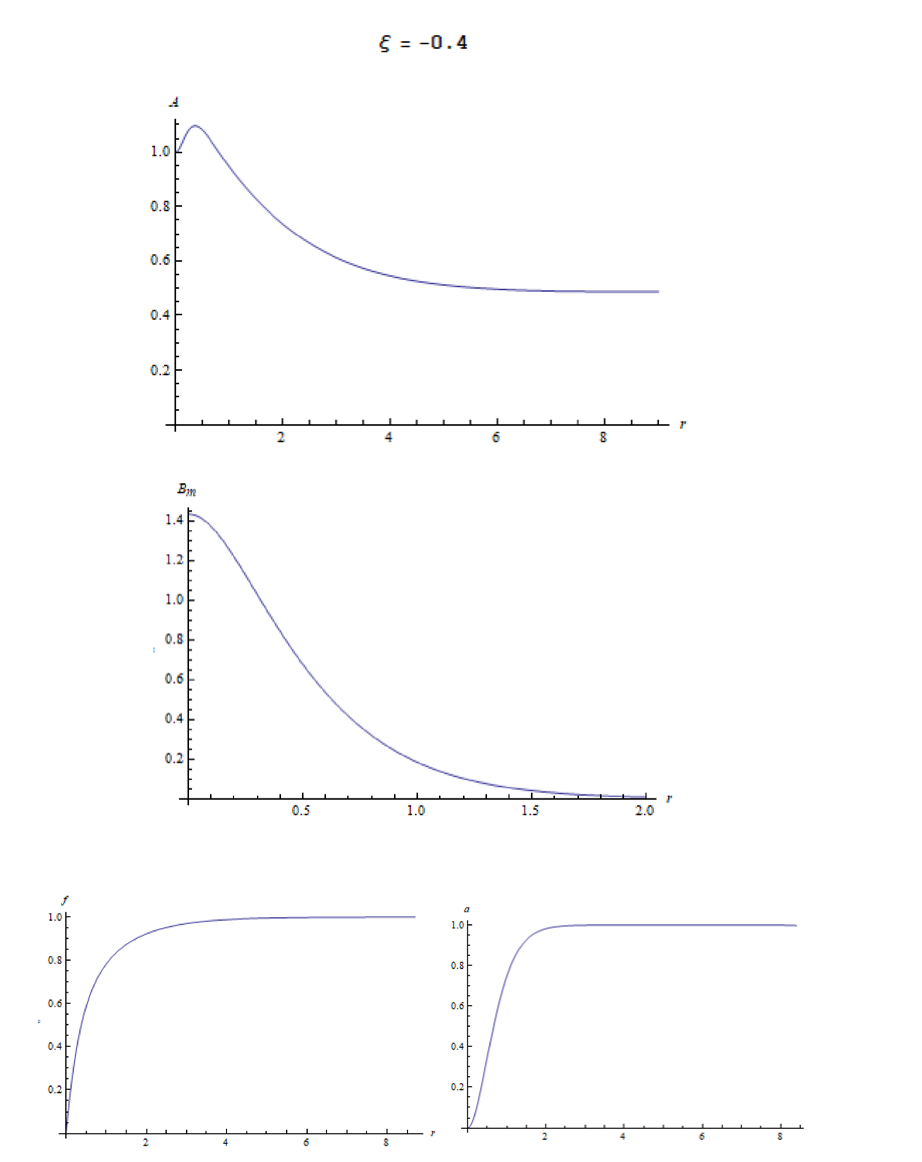}
		\caption{Flat case $\xi=-0.4$. We plot the metric $A$, the magnetic field $B_m$ and the scalar $f$ and gauge field $a$. Since the VEV of the scalar field is always unity, it plateaus at unity regardless of the value of $\xi$. The gauge field $a$ always plateaus at unity also since $n=1$ for all $\xi$. We therefore show the scalar and gauge field profile here but not in subsequent figures, since they are roughly similar. The metric profile plateaus at $D=0.488$ which yields a deficit angle of $1.894$ rad, the largest deficit angle in our sample but not the one with the highest mass (see table \ref{Table2}). The magnetic field peaks at $1.43$, which is the highest peak in our sample. This implies that it extends the least (falls off fastest) since the magnetic flux  remains constant at $\Phi=2 \pi n/e=2.0944$ to within three or four decimal places.}     
	\label{Flat1}
\end{figure}
  \begin{figure}[!htb]
		\includegraphics[scale=0.8]{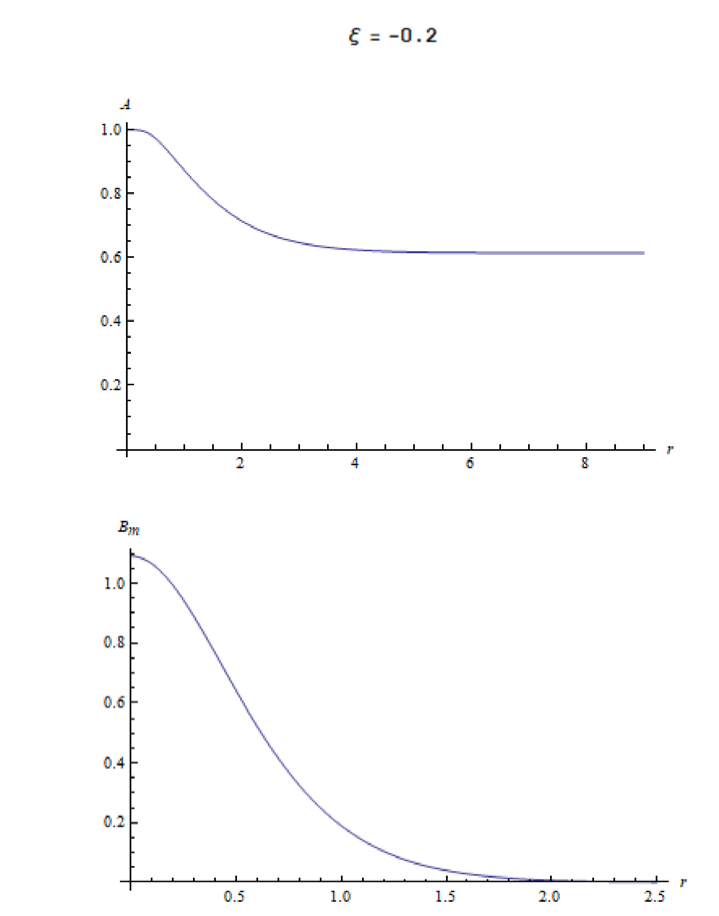}
		\caption{Flat case $\xi=-0.2$. The metric $A$ plateaus at $D=0.615$ yielding a  deficit angle $1.356$ rad, the second largest deficit angle in our sample. The magnetic field peaks at $1.092$, the second largest peak in our sample.}     
	\label{Flat2}
\end{figure}\begin{figure}[!htb]
		\includegraphics[scale=0.8]{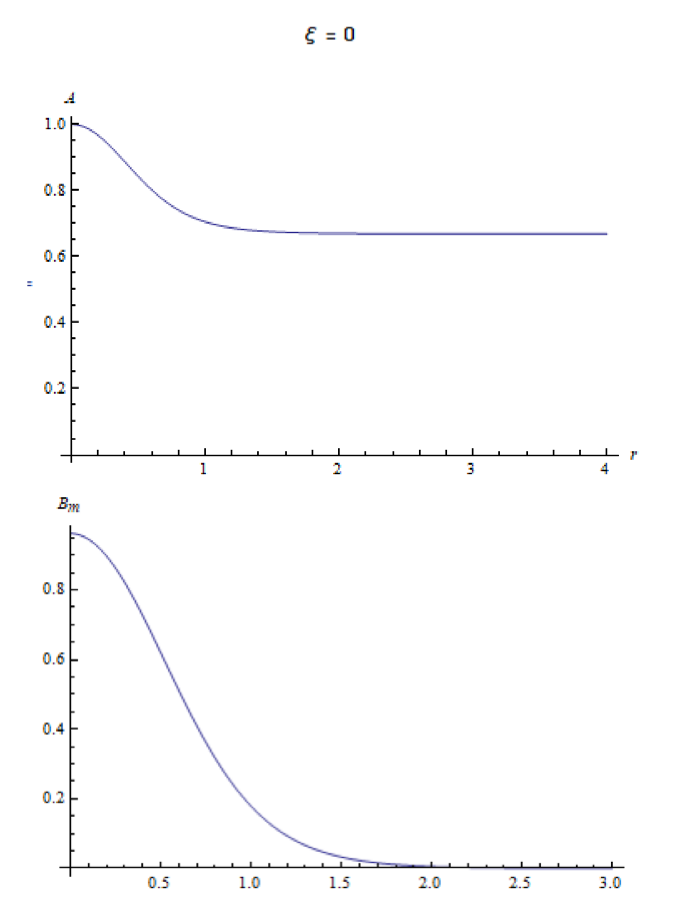}
		\caption{Flat case $\xi=0$. The non-minimal coupling is turned off here. The metric $A$ plateaus at $D=0.668$ yielding a deficit angle of $1.148$. The magnetic field peaks at $0.962$ and is less than the previous case. }     
	\label{Flat3}
\end{figure}\begin{figure}[!htb]
		\includegraphics[scale=0.8]{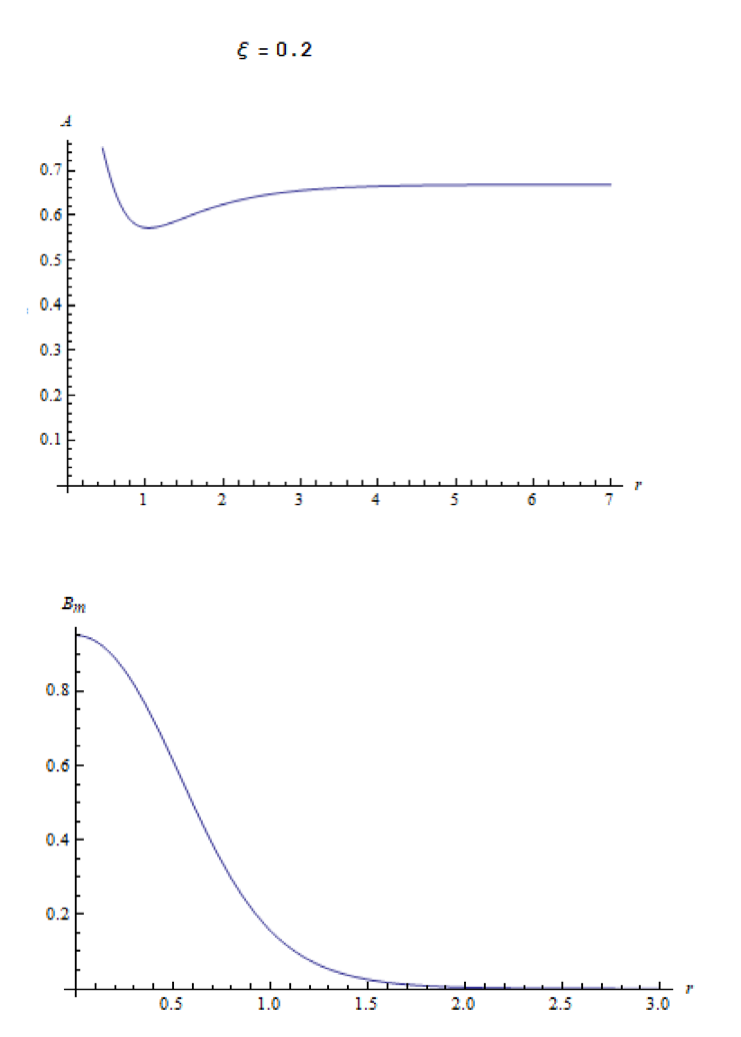}
		\caption{Flat case $\xi=0.2$. The metric $A$ plateaus at $D=0.668$, the same value as the previous case. It therefore also has a deficit angle of $1.148$. It has a peak magnetic field of $0.950$ which is less than the previous case. Up to here so far, there has been a trend: the peak of the magnetic field has monotonically decreased and the deficit angle has decreased or remained the same.}     
	\label{Flat4}
\end{figure}\begin{figure}[!htb]
		\includegraphics[scale=0.8]{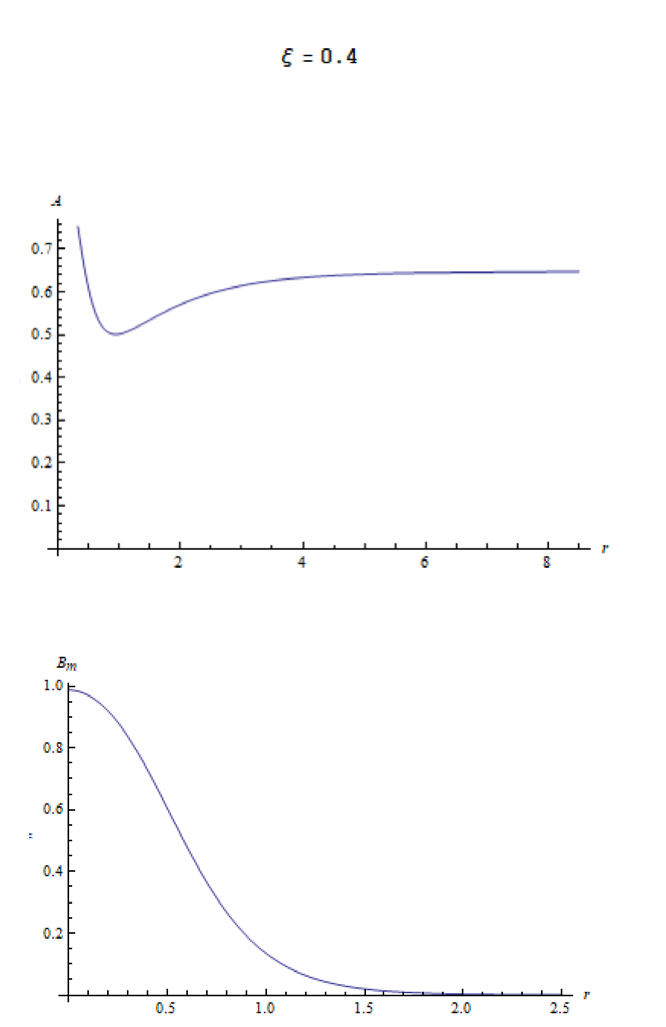}
		\caption{Flat case $\xi=0.4$. This case departs from the above decreasing trend. The metric plateaus at $D=0.648$ yielding a deficit angle of $1.225$ rad and a peak magnetic field of $0.987$: both are greater than in the previous two cases.}     
	\label{Flat5}
\end{figure}   

\begin{table}[b]
	\centering
		\includegraphics[scale=1]{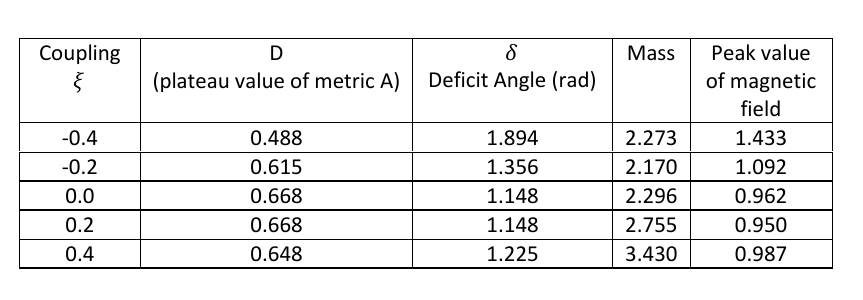}
		\caption{The most important thing about this table is that the deficit angle is not proportional to the mass. Compare the first and last row. At $\xi=-0.4$ one has the largest deficit angle of $1.894$ rad with a mass of $2.273$ whereas at $\xi=+0.4$ the mass is significantly higher at $3.430$ and yet it has a much smaller deficit angle of $1.225$ rad. With the non-minimal coupling term present, the ratio of mass to deficit angle is not constant but depends on $\xi$ (see body of text).}     
	\label{Table2}
\end{table} 

\section{Conclusion}
In this paper, we studied the effects of the non-minimal coupling term 
$\xi\,R\,|\phi|^2$ on a vortex under Einstein gravity in an AdS$_3$ and flat (conical) background. In the case of AdS$_3$, this led to the emergence of a critical coupling $\xi_c$ where the VEV of the scalar field is zero for $\xi$ at or above $\xi_c$ but is non-zero when $\xi$ crosses below $\xi_c$. For the values of our parameters, $\xi_c$ was equal to $2/21\approx 0.0952$. We presented our numerical results in plots and tables for nine values of $\xi$. Our plot of the numerically obtained VEV versus $\xi$ was in accord with the theoretical expectation that the slope has a discontinuity and diverges at the critical coupling $\xi_c$. For $\xi$ near $\xi_c$, we verified numerically that the VEV indeed behaved according to the power law $|\xi-\xi_c|^{1/2}$. These results confirmed the idea that the critical coupling $\xi_c$ acts like the analog of the critical temperature $T_c$ in GL mean-field theory. In that theory, the order parameter is zero at or above $T_c$ and is non-zero below $T_c$ and behaves according to the power law $|T-T_c|^{1/2}$. The plot of the order parameter versus temperature $T$ also shows a discontinuity and divergence in the slope near $T_c$. Numerical results of the 
``extension" of the scalar field (core region of the vortex) show that it increases monotonically as $\xi$ increases, with a dramatic increase near $\xi_c$. We showed analytically that it is expected to diverge at the critical coupling and this is analogous to the divergence of the coherence length in GL mean-field theory as one approaches the critical temperature.    

In asymptotically flat (conical) spacetime, we considered five values of $\xi$ and remarkably, found that higher masses did not necessarily lead to a higher deficit angle as one might naively expect. The reason for this is that, when a non-minimal coupling term is present, the  ratio of mass to deficit angle is no longer constant but depends on the coupling $\xi$. This can lead to cases where a higher mass has a smaller deficit angle than a smaller mass as our data clearly showed.    

If $\xi_c$ acts as the analog to $T_c$ in GL mean-field theory, this naturally raises the question, ``Is the non-minimally coupled vortex a thermodynamic system at non-zero temperature?". The answer is clearly no. The Nielsen-Olesen vortex without gravity constitutes a static classical field configuration which is at zero temperature and has zero entropy. The zero temperature agrees with the fact that the fields have no average kinetic energy and the zero entropy is in accord with the fact we know everything about the field's configuration throughout spacetime; we are not ignorant of its configuration at any time and no information is hidden from us. The zero entropy is of course consistent with the zero temperature. When gravity is included, this can change only if the vortex acquires an event horizon. However, our gravitating vortex solutions are non-singular static solutions with no event horizon. The temperature and entropy are again zero and as before, the metric field, as well as the scalar and gauge field, are static throughout all of spacetime. In contrast, the BTZ black hole \cite{BTZ1,BTZ2} has a non-zero temperature and entropy as it has an event horizon (for simplicity, assume no angular momentum or electric charge, only mass with a single horizon).  Note that the BTZ spacetime has a timelike Killing vector outside the event horizon but like the Schwarzschild black hole in $3+1$ dimensions, it has no timelike Killing vector inside the event horizon \cite{Extremal}. This implies that there is no coordinate transformation that can put the metric in static form inside the event horizon so that an outside observer is ignorant of the metric configuration inside at any particular time. Simply put, information is hidden from us behind the event horizon \cite{Bekenstein}. Note that in contrast, our non-singular gravitating static vortex has a timelike Killing vector throughout spacetime and no information is hidden from us (see also \cite{Carroll2, Horowitz, Teitelboim} for a related discussion). 

The vortex actually constitutes a classical solution in quantum field theory (QFT) \cite{Weinberg}. The vortex cannot be obtained from perturbative QFT as it is a non-perturbative solution. It turns out that since the size of the vortex is much larger than its compton wavelength, the classical non-perturbative solution constitutes a valid solution to the QFT (i.e. a very good first approximation) \cite{Weinberg}. Perturbation theory can then be used to obtain one-loop quantum corrections to the vortex by quantizing about the classical configuration. In particular, quantum fluctuations of the scalar field will change the nature of the potential as there will now be logarithmic terms besides the usual terms \cite{Zee, ArielNoah}. The critical exponent of $1/2$ will therefore change as a consequence of these quantum corrections. So an interesting and pertinent problem to solve for the future is to determine the critical exponent of the non-minimally coupled vortex in an AdS$_3$ background after quantum corrections. This would be a considerably more complicated calculation than say the quantization about the $1+1$-dimensional kink in Minkowski spacetime \cite{Weinberg} as we have one extra spatial dimension and a curved space background.

\pagebreak
\begin{appendices}
\numberwithin{equation}{section}
\setcounter{equation}{0}

\section{Derivation of the VEV $v_{eff}$ and cosmological constant $\Lambda_{eff}$}
\numberwithin{equation}{section}
\setcounter{equation}{0}
In this appendix we derive the expressions for $v_{eff}$ and $\Lambda_{eff}$ given by equations \reff{Veff} and \reff{Leff} respectively. We start by rewriting the equations \reff{Vev1} and \reff{Cosmo1} where $v_{eff}$ and $\Lambda_{eff}$ are expressed in terms of each other: 
\begin{align}
v_{eff}^2= v^2 + \dfrac{12 \,\xi \,\Lambda_{eff}}{\lambda}\,.
\label{Vev2}
\end{align}
\begin{align}
\alpha (R-2\,\Lambda) +\xi \,R\, v_{eff}^2- \dfrac{\lambda}{4}(v_{eff}^2-v^2)^2=(\alpha + \xi \, v_{eff}^2)(R- 2\,\Lambda_{eff})\,.
\label{Cosmo2}
\end{align}
We first substitute the asymptotic value of the Ricci scalar, $R= 6 \,\Lambda_{eff}$, into \reff{Cosmo2} which yields
\begin{align}
\Lambda_{eff}=\dfrac{\alpha \,\Lambda + \dfrac{\lambda}{8}(v_{eff}^2-v^2)^2}{\alpha + \xi\, v_{eff}^2}\,.
\label{Cosmo3}
\end{align} 
Substituting \reff{Cosmo3} into \reff{Vev2} yields a quadratic equation for $v_{eff}^2$:
\begin{align}
\lambda \xi \,(v_{eff}^2)^2 -2\,\lambda (\alpha+2 v^2 \xi)\, v_{eff}^2 +2\,v^2 \alpha \lambda+3\, v^4 \lambda \xi + 24 \,\alpha \Lambda\xi =0\,.
\label{Veff3}
\end{align}
This yields the following two possible solutions for $v_{eff}^2$ (which we label I and II):
\begin{align}
\text{I:}\quad 2 v^2+\frac{\alpha }{\xi }-\frac{\sqrt{\alpha ^2+2 v^2 \alpha  \xi +v^4 \xi ^2-24 \alpha  \Lambda  \xi ^2/\lambda }}{\xi }
\label{I}
\end{align}
\begin{align}
\text{II:} \quad 2 v^2+\frac{\alpha }{\xi }+\frac{\sqrt{\alpha ^2 +2 v^2 \alpha \xi +v^4 \xi ^2-24 \alpha   \Lambda  \xi ^2/\lambda}}{ \xi }
\end{align}
However, only the first solution satisfies the requirement that $v_{eff}$ is equal to $v$ in the limit $\xi \to 0$. The second solution yields $\infty$ in that limit and must be disregarded. Taking the positive of the square root of the first solution yields the quoted result \reff{Veff} for $v_{eff}$:
\begin{align}
v_{eff}=\Bigg[2 v^2 + \frac{\alpha}{\xi} - \frac{\sqrt{\alpha^2 + 
  2 \,v^2\,\alpha \,\xi + v^4\,\xi^2 - 24 \,\alpha\,\Lambda\,\xi^2/\lambda}}{\xi}\Bigg]^{1/2}
\label{Veff4}
\end{align}  
Substituting the above solution \reff{Veff4} into \reff{Cosmo3} yields the quoted result \reff{Leff} for $\Lambda_{eff}$:
\begin{align}
\Lambda_{eff}=\dfrac{\lambda}{12\,\xi^2}\Big(\alpha + v^2\,\xi -\sqrt{\alpha^2+ v^4\,\xi^2 +2\,v^2\,\alpha\,\xi-24\,\alpha\,\Lambda\,\xi^2/\lambda}\Big)\,.
\label{Leff4}
\end{align}

\section{Full equations of motion}
\numberwithin{equation}{section}
\setcounter{equation}{0}
The three equations of motion quoted in the text are \reff{EOMB2},\reff{EOMf2} and \reff{EOMa2}. Equation \reff{EOMa2} contains the function $W(r)=B'/B$ and equation \reff{EOMf2}contains $W$ and its derivative $W'$. We can extract $W$ from \reff{EOMA} and this yields
\begin{align} 
W=&\dfrac{1}{4 e^2 r A \left(\alpha +\xi  f^{\,2}+2 r \xi  f f'\right)}
\Big(-e^2 r^2 \left(v^4 \lambda +8 \alpha  \Lambda \right)-2 e^2 (n^2-r^2 v^2 \lambda \nonumber \\&\quad\quad-2\, n \,a+a^2) f^2-e^2 r^2 \lambda  f^4-16 e^2 r \xi A f f'+2 A (a'^{\,2}+ e^2 r^2 f'^{\,2})\Big)\,.                
\label{W}
\end{align}
Substituting the above expression for $W$ (as well as its derivative) back into \reff{EOMf2} and \reff{EOMa2} and keeping \reff{EOMB2} the same yields three equations of motion that have no dependence on the function $B$. The full three equations are:
\begin{align}
&e^2 r^2 \lambda f^4+e^2 r \,(r v^4 \lambda+8 r \alpha \Lambda+4 \alpha A')+2 e^2 f^2 (n^2-r^2 v^2 \lambda-2\, n \,a+a^2+2 r \xi A')\nonumber\\&\quad\quad+2 A \,(a'^{\,2}+e^2 r^2 (1+8 \xi) f'^{\,2})+8 e^2 r \xi f \,\big(r A' f'+2 A \,(f'+r f'')\big)=0\,.\label{EOMB4}
\end{align}
\begin{align}
&-2 r^2 \lambda  f^3-2 f \left(n^2-r^2 v^2 \lambda -2 n a+a^2+2 r \xi  A'\right)+r (r A' f'+2 A (f'+r f''))\nonumber\\&+\dfrac{1}{8 e^4 A (\alpha +\xi  f (f+2 r f'))^2}\Big(\xi  f (e^2 r^2 (v^4 \lambda +8 \alpha  \Lambda )-2 A a'^{\,2}\nonumber\\&+e^2 (2 (n^2-r^2 v^2 \lambda -2 n a+a^2) f^2+r^2 \lambda  f^4+16 r \xi  A f f'-2 r^2 A f'^{\,2}))^2\Big)\nonumber\\&
+\dfrac{r}{4 e^4 A (\alpha +\xi  f (f+2 r f'))^2}\,\Bigg(2 e^2 \xi  f A' (\alpha +\xi  f (f+2 r f')) (e^2 r^2 (v^4 \lambda +8 \alpha  \Lambda )-2 A a'^{\,2}\nonumber\\&+e^2 (2 (n^2-r^2 v^2 \lambda -2 n a+a^2) f^2+r^2 \lambda  f^4 +16 r \xi  A f f'-2 r^2 A f'^{\,2}))\nonumber\\&-e^2 A f' (\alpha +\xi  f (f+2 r f'))(e^2 r^2 (v^4 \lambda +8 \alpha  \Lambda )-2 A a'^{\,2}\nonumber\\&+e^2 (2 (n^2-r^2 v^2 \lambda -2 n a+a^2) f^2+r^2 \lambda  f^4+16 r \xi  A f f'-2 r^2 A f'^{\,2}))\nonumber\\&+\frac{1}{r}\xi  f \bigg (-e^4 (r^2 (v^4 \lambda +8 \alpha  \Lambda )+2 (n^2-r^2 v^2 \lambda -2 n a+a^2) f^2+r^2 \lambda  f^4) \nonumber\\&(r^2 \lambda  f^4+r (r v^4 \lambda +8 r \alpha  \Lambda +4 \alpha  A')+2 f^2 (n^2-r^2 v^2 \lambda -2 n a+a^2+2 r \xi  A')+8 r^2 \xi  f A' f')\nonumber\\&-4 e^2 A (-2 e^2 r^2 \lambda  \xi  f^6-r^2 (v^4 \lambda +8 \alpha  \Lambda ) (a'^{\,2}+e^2 (2 \alpha +r^2 (1-2 \xi ) f'^{\,2}))\nonumber\\&-r f^4 (4 e^2 \xi  (-n+a) a'+r \lambda  a'^{\,2}+e^2 r \lambda  (2 (\alpha -2 v^2 \xi )+r^2 (1+6 \xi ) f'^{\,2}))\nonumber\\&-2 f^2 (2 e^2 r \alpha  (-n+a) a'+(n^2-r^2 v^2 \lambda -2 n a+a^2) a'^{\,2}+e^2 r^2 (-2 v^2 \alpha  \lambda +v^4 \lambda  \xi +8 \alpha  \Lambda  \xi \nonumber\\&+(1+2 \xi ) (n^2-r^2 v^2 \lambda -2 n a+a^2) f'^{\,2}))+2 e^2 r^3 \lambda  \xi  f^5 (2 f'+r f'')\nonumber\\&+2 e^2 r f (2 (-n^2 \alpha +r^2 (v^2 \alpha  \lambda +2 v^4 \lambda  \xi +16 \alpha  \Lambda  \xi )+\alpha  (2 n-a) a) f'+r^3 (v^4 \lambda +8 \alpha  \Lambda ) \xi  f'')\nonumber\\&+4 e^2 r f^3 (-(-5 n^2 \xi +r^2 \lambda  (\alpha +3 v^2 \xi )+5 \xi  (2 n-a) a+2 r \xi  (-n+a) a') f'\nonumber\\&+r \xi  (n^2-r^2 v^2 \lambda -2 n a+a^2) f''))-4 A^2 \Big(a'^{\,4}-16 e^4 r \xi  f (\alpha +\xi  f^2) f'\nonumber\\&+4 e^2 r a' (\alpha +\xi  f (f+2 r f')) a''-2 e^2 r a'^{\,2} (r (-1+2 \xi ) f'^{\,2}+2 \xi  f (6 f'+r f''))\nonumber\\&+e^4 r^2 \big(-16 r \xi  f f'^{\,3}+r^2 (1-4 \xi ) f'^{\,4}-16 \xi  f (\alpha +\xi  f^2) f''\nonumber\\&+4 r (\alpha +\xi  f^2) f' f''+4 f'^{\,2} (\alpha -4 \alpha  \xi +\xi  f (f+20 \xi  f+r^2 f''))\big)\Big)\bigg)\Bigg)=0\,.\label{EOMf4}
\end{align} 
\begin{align}
&2 e^2 r (n-a) f^2-2 A a'+r a' A'+2 r A a''\nonumber\\&+\dfrac{a'}{4 e^2 \left(\alpha +\xi  f^2+2 r \xi  f f'\right)} \Big(-e^2 r^2 \left(v^4 \lambda +8 \alpha  \Lambda \right)\nonumber\\&-2 e^2 \left(n^2-r^2 v^2 \lambda -2 n a+a^2\right) f^2-e^2 r^2 \lambda  f^4-16 e^2 r \xi  A f f'+2 A \left(a'^{\,2}+e^2 r^2 f'^{\,2}\right)\Big)=0\,.
\label{EOMa4}
\end{align} 
The above three equations are those we solve numerically. 
\end{appendices}

\section*{Acknowledgments}
A.E. acknowledges support from a discovery grant of the National Science and Engineering Research Council of Canada (NSERC).

\end{document}